\documentclass[aps,%
               prl,%
               reprint,%
               showpacs,%
               superscriptaddress,%
               longbibliography,%
               nofootinbib%
               ]{revtex4-2}
\usepackage{svg}
\usepackage[utf8]{inputenc}
\usepackage{graphicx}
\usepackage{amsfonts}
\usepackage{amssymb}
\usepackage{amsmath}
\usepackage{mathtools}
\usepackage{bm}
\usepackage{dsfont}
\usepackage{braket}
\usepackage{tikz}
\usepackage{txfonts}
\usepackage{float} 
\usepackage[pdfusetitle,%
            bookmarks=true,%
            colorlinks,%
            linkcolor=blue,%
            citecolor=blue,%
            urlcolor=blue%
            ]{hyperref}

\newcommand{\fig}[1]{Fig.\thinspace{}\ref{#1}}
\newcommand{\fc}[1]{({#1})}
\newcommand{\figc}[2]{Fig.\thinspace{}\ref{#1}\thinspace{}\fc{#2}}
  
\newcommand{\subfigref}[2]{\hyperref[fig:#1]{\ref*{fig:#1}(#2)}}

\usepackage{lipsum}

\begin{document}

\def\papertitle{{Theory of Angle Resolved Photoemission Spectroscopy of Altermagnetic Mott Insulators}}

\def\tum{{Technical University of Munich, TUM School of Natural Sciences, Physics Department, 85748 Garching, Germany}}
\def\mcqst{{Munich Center for Quantum Science and Technology (MCQST), Schellingstr. 4, 80799 M{\"u}nchen, Germany}}
\def\iis{{Indian Institute of Science, Bangalore, 560012, India}}

\newcommand{\TUM}{\affiliation{\tum}}
\newcommand{\MCQST}{\affiliation{\mcqst}}
\newcommand{\IIS}{\affiliation{\iis}}

\title{\papertitle}
\author{Lorenzo Lanzini}
\TUM
\MCQST

\author{Purnendu Das}
\IIS
\TUM
\MCQST
\author{Michael Knap}
\TUM
\MCQST
\date{\today}

\begin{abstract}
\noindent Altermagnetism has emerged as an unconventional form of collinear magnetism with spatial rotational symmetries, that give rise to strongly spin-split bands despite of an underlying fully-compensated antiferromagnetic order. Here, we develop a theory for the Angle Resolved Photoemission Spectroscopy (ARPES) response of altermagnetic Mott insulators. Crucially, the spectrum does not simply reflect the non-interacting band structure, but instead a magnetic polaron is formed at low energies, that can be interpreted as a spinon-holon bound state. We develop a spinon-holon parton theory and predict a renormalized bandwidth that we confirm by tensor network simulations. We analyze the characteristic spin-split spectrum and identify a spin-dependent spectral weight of the magnetic polaron, resulting from the altermagnetic symmetry. Our work paves the way for a systematic study of doping effects and correlation phenomena in altermagnetic Mott insulators.
\end{abstract}

\maketitle

\textbf{\textit{Introduction.---}}Altermagnets are collinear magnets, that in some respect are distinct from ferromagnets and from conventional antiferromagnets~\cite{Smejkal2022a, Smejkal2022b}. Although altermagnets have zero net magnetization resembling antiferromagnets, an underlying rotational symmetry between the sublattices gives rise to spin-split bands~\cite{Pomeranchuk1958,wu2007fermi,Ahn2019,Hayami_2019,Smejkal2020crystal,Hayami2020,mazin2021prediction,Smejkal2022b,Shao2021,smejkal2022giant,Smejkal2022a,Mazin2022editorial, Maier2023, Leeb2024, Das2024, Ferrari2024, D'Ornellas2025, he2025, wu2025, vijayvargia2025,  jaeschkeubiergo2025, Sato2024, Liu_2025, Chang_2025, Parshukov2025, DelRe_2025, Regmi2025, devita2025, gong2025, Zhu2025, wiedmann2025,peces2025,ni2025, sicheler2025}. The spin splitting can be directly probed by Angle Resolved Photoemission Spectroscopy (ARPES)~\cite{Sobota2021}. Recently (spin-resolved) ARPES and other spectroscopic probes have been measured for a range of altermagnetic candidate materials 
~\cite{Lee2024, Krempaský2024, Osumi2024, Reimers2024, Ding2024, li2024, Fedchenko2024, Yang2025}.
{For instance, a very strong spin-split band structure has been observed in the weak insulator MnTe~\cite{Gonzalez_Betancourt2023, Lee2024, Amin2024, gray2024, Krempaský2024, Hariki2024, Osumi2024, Bey2025}. Many altermagnetic materials have also been argued to be strongly correlated Mott insulators~\cite{Cuono_2023, GUO_2023, Maznichenko_2024}, where the Coulomb repulsion causes strong localization of electrons. Among them is the recently discovered $\mathrm{La_2O_3Mn_2Se_2}$, a Mott insulating layered material exhibiting d-wave altermagnetism~\cite{Wei_2025,Garcia-Gassull2026}.} 

The APRES response should be interpreted as a measure of the dynamics of charge dopants injected into the altermagnet. In general, these dopants do not reflect the bare band structure of the material, but instead a magnetic polaron is formed due to interaction with the magnetic environment~\cite{Schmitt-Rink1988, Kane1989, Sachdev1989, StrackVollhardt1992, LoganStumpf1998, Sangiovanni2006, Taranto2012}. For conventional antiferromagnets, an intuitive picture capturing the magnetic polaron formation is obtained from a parton description in which the doped hole splits into a spinon and a holon~\cite{StrackVollhardt1992, B_ran_1996,Laughlin1997, Grusdt2018a, Grusdt2019, Bohrdt2020, Bohrdt_2020_b, Wrzosek_2021}. In this picture, the magnetic polaron is a quasiparticle that is a bound state of the fast holon and the slow spinon. Despite the wealth of recent APRES experiments, the consequences of the magnetic polaron formation in altermagnetic insulators has not been investigated thus far. 

\begin{figure}[H]
    \centering
    \hspace*{-2mm}
    \includegraphics[scale=0.92]{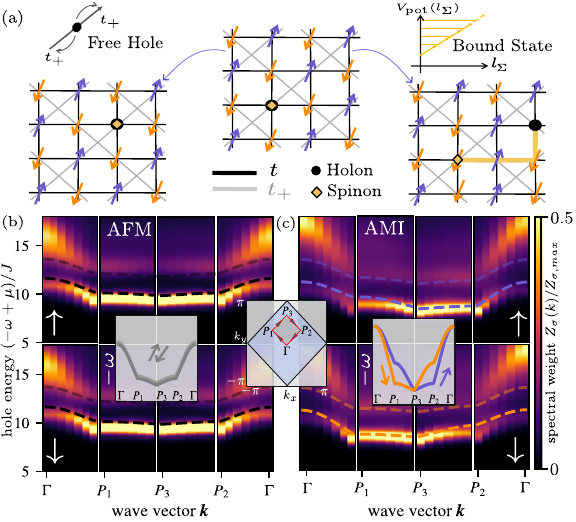}
    \caption{\textbf{The altermagnetic polaron.} (a) Center: A hole doped into an altermagnetic Mott insulator with Néel order can be interpreted as tightly-bound spinon (diamond) and holon (circle). Two main processes contribute to the dynamics. Left: anisotropic diagonal hopping $t_+$ of the spinon-holon composite that retains the Néel order; Right: nearest-neighbor hopping $t$ that displaces the magnetic background and thus generates an effective confining potential for the spinon and holon. Spin-resolved ARPES response (top spin-$\uparrow$, bottom spin-$\downarrow$), in (b) the antiferromagnet (AFM) and (c) the altermagnetic Mott insulator (AMI), obtained with tensor network simulations for $t_+/t = 0.4, t/J = 3$ (color plot) compared with the parton theory (dashed lines). Both cases show a quasiparticle peak---the magnetic polaron---at low energies. Such peak is characterized by a strongly reduced bandwidth ($W \sim \mathcal{O}(J)$ respect to the bandwidth at the mean-field level $W \sim \mathcal{O}(t)$). While for the AFM the spectra are spin-symmetric, \textit{c.f.} inset of (b), for the AMI they are anisotropic and are related by the altermagnetic $\pi/2$-symmetry of the model, \textit{c.f.} inset of (c). In both AFM and AMI spectra, a higher energy vibrational peak of the magnetic polaron is observed (faint dashed line).}
    \label{fig:1}
\end{figure}

In this work, we determine the spin-resolved ARPES response of a $d$-wave altermagnetic Mott insulator~\cite{Das2024} using tensor network methods and develop a parton theory~\cite{Grusdt2018a} for the magnetic polaron formation; see \fig{fig:1}. Our theory predicts a strong renormalization of the quasi-particle bandwidth due to interactions with the antiferromagnetic background; however, in a more complex way compared to simple square-lattice antiferromagnets.
 We determine the characteristic spin-splitting of the bands and analyze the spin-anisotropic spectral weight of the magnetic polaron, that we argue are key signatures of altermagnetism.
 
\textbf{\textit{The altermagnetic Mott insulator.---}}We consider an altermagnetic Hubbard model on a square lattice $\hat H = - \sum_{i,j, \sigma} t_{ij} (c_{i \sigma}^\dagger c_{j \sigma} + \text{h.c}) + U \sum_{i} n_{i \uparrow}n_{i \downarrow}$, where $c_{i\sigma}$ ($c_{i\sigma}^{\dagger}$) are the fermionic annihilation (creation) operators \cite{Das2024}. The hopping matrix elements are $t_{ij}=t$ for nearest-neighbors, while $t_{ij}=t_+$ in the $\bm{a}_1=(1,1)$ direction for the A sublattice and in the $\bm{a}_2=(-1, 1)$ direction for the B sublattice, and zero for all the other bonds; see \figc{fig:1}{a}. This model has an altermagnetic $\pi/2$-symmetry ($d$-wave symmetry), expected to capture e.g. the $\mathrm{Mn_2O}$ layers in $\mathrm{La_2O_3Mn_2Se_2}$ \cite{Wei_2025, Garcia-Gassull2026}. 

In the limit of strong interactions $(U \gg t)$, we project out double occupancies using a Schrieffer-Wolff transformation to obtain an effective $t-J$ model
\begin{equation}\label{eq: alt tj model eq}
    \mathcal{H} =  \sum_{ij, \sigma} - t_{ij}(\mathcal{P}_{GW}c^{\dagger}_{i\sigma}c_{j\sigma}\mathcal{P}_{GW} + h.c) + J_{ij}(\bm{S}_i\cdot\bm{S}_j-\frac{n_in_j}{4})
\end{equation}
where $J_{ij} = 4t_{ij}^2/U$ are the spin-exchange couplings and $\mathcal{P}_{GW}$ is a Gutzwiller projection to the single occupied local basis states. Third-order terms involving three sites on a triangle vanish for real hopping \cite{Motrunich2006}. The spin excitations of insulating altermagnets in the absence of doping have been analyzed recently, see e.g. Refs.~\cite{Costa2025, kaushal2024}. Here, we focus on the single-particle spectral function that is measured in ARPES experiments.

\textbf{\textit{Single-particle spectral function.---}}In order to numerically compute the single-particle spectral function for large systems, we first perform tensor network simulations to determine the ground state $|\Psi_0\rangle$ of model \eqref{eq: alt tj model eq} at half filling. We add a weak sublattice potential to break the SU(2) symmetry of the model and to spontaneously form Néel order. We represent the wave function as a matrix product state (MPS) on elongated cylinder geometries and variationally optimize the ground state with the \textit{Density Matrix Renormalization Group} algorithm~\cite{white} using the TeNPy package \cite{Hauschild2018}. We then compute the time-dependent correlation function $\langle \Psi_0| c^{\dagger}_{\bm{r}_b\sigma}(\tau)c_{\bm{0}_a\sigma}|\Psi_0\rangle$, where the indices $a,b$ of the real space coordinates indicate the position in the two-site unit cell and $\tau$ is real time. From this correlation function, we obtain the spin-resolved spectral function
\begin{equation}
    A_{\sigma}(\mathbf{k},\omega) = \sum_{a,b} \int \mathrm{d} \mathbf{r}_b  \mathrm{d} \tau e^{i(\tau \omega- \mathbf{k} \mathbf{r}_b-k_y(b-a))} \langle \Psi_0| c^{\dagger}_{\bm{r}_b\sigma}(\tau)c_{\bm{0}_a\sigma}|\Psi_0\rangle,
    \label{eq::A}
\end{equation}
measured with spin-resolved ARPES.

\begin{figure}
    \centering
    
    \includegraphics[scale=0.95]{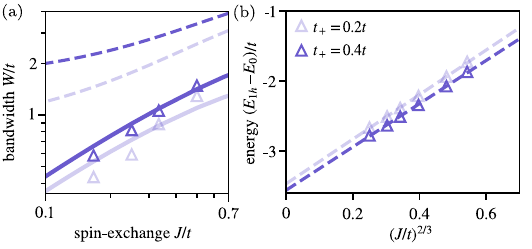}
    \caption{\textbf{Signatures of the polaron formation.} 
    (a) Bandwidth of the altermagnetic polaron extracted from the spectral function, that we compute with tensor networks (triangles), compared with the parton theory (solid lines). The bandwidth $W$ of the polaron is strongly renormalized compared to the Hartree-Fock prediction (dashed lines).
    (b) Energy difference of the single-hole and the undoped ground state (triangles), which we find to scale as $\sim (J/t)^{2/3}$ (dashed line), suggesting the formation of a linear confinement potential between the holon and the spinon. 
    }
    \label{fig:fig2}
\end{figure}

We now determine the spin-resolved spectral function both for the square-lattice antiferromagnet $t=3J, t_+=0$ and the altermagnet $t=3J, t_+=0.4t$; see Fig.~\ref{fig:1}(b,c). We have introduced a small staggering field \( B = 5 \cdot 10^{-3} J\) to the Hamiltonian to break the spin symmetry in the z-direction, resulting in a staggered magnetization \( m \simeq 0.25 \), which is half of the saturated Néel order. {This reduction is due to the strong quantum fluctuations that in 2D reduce the Néel order parameter. It is consistent with a linear spin-wave analysis of altermagnetic Heisenberg model, see supplemental material~\cite{supp}}. We consider cylinder geometries elongated in the $\bm{a}_1$($\bm{a}_2$) direction, leading to a fine resolution of the shown Brillouin zone cuts. {We determine the chemical potential by calculating the DMRG energies with $N+1$ and $N-1$ electrons, $\mu_{hf} = [{E(N+1)-E(N-1)}]/{2} $, such that the Mott gap can be directly read off from the spectra~\cite{Kadow_2022}.} The numerical results obtained from the tensor network simulations are checked to be converged with bond dimension of the matrix product state, which is the control parameter for the accuracy of the simulations; see supplemental material for a detailed analysis~\cite{supp}. Comparing the antiferromagnet and the altermagnet, both show a well-defined quasiparticle peak at low energies, indicating the formation of the magnetic polaron. We further find the minimum of the dispersion both for the altermagnet and the antiferromagnet at the nodal point $Q_2=(\pi/2, \pi/2)$.

Crucially, we also identify key differences between the antiferromagnet and the altermagnet. First, the antiferromagnet has spin degenerate bands while the altermagnet does not. In particular, the bands in the altermagnet are spin split and the spectra are related by a $\pi/2$-rotation expected for the $d$-wave altermagnet of Eq.~\eqref{eq: alt tj model eq}. Second, in the altermagnet the spectral weight is not uniform for the two spin orientations, which we propse as an additional experimental signature of altermagnetic Mott insulators. 

We now determine the single-particle spectral function for various values of the spin exchange $J$ and diagonal hoppings $t_+$ and extract the bandwidth $W$ of the magnetic polaron; see Fig.~\ref{fig:fig2}(a). The bandwidth $W$ increases with spin-exchange $J$ and with diagonal hopping $t_+$, however, it is strongly reduced compared to the Hartree-Fock prediction, hinting at a strong renormalization of the doped hole. To gain further understanding of the structure of the state, we compute the energy difference of the ground state at half filling and the one with a doped hole. We find that this energy difference scales as $\Delta E/t \propto (J/t)^{2/3}$, see \figc{fig:fig2}{b}, reminiscent of the polaron formation in the conventional square-lattice antiferromagnet~\cite{Brinkman1970, Manousakis2007, Grusdt2019}.

\textbf{\textit{Microscopic Spinon-Holon Parton Theory.---}}In our tensor network simulations we observed a strong renormalization of the magnetic polaron by the magnetic background. In order to interpret this process, we develop a parton theory \cite{Grusdt2019,Bohrdt2020}, in which the fermionic hole fractionalizes into a heavy spinon that carries the spin quantum number and a light bosonic holon that carries the charge; see \figc{fig:1}{a}. Qualitatively, two distinct processes can arise for a square lattice antiferromagnet with both nearest and next-nearest neighbor hopping. On the one hand, the diagonal hopping leaves the Néel ordered background invariant and the spinon and holon remain tightly bound (left panel). On the other hand, the nearest-neighbor hopping leads to a separation of the light holon and the heavy spinon, creating geometric strings of displaced spins (right panel).
The properties of the displaced strings can be evaluated using a Born-Oppenheimer type approximation, in which the hole does not change the quantum state or entanglement but only displaces spins, which is referred to as \textit{Frozen Spin Approximation} (FSA) \cite{Chiu2019,Grusdt2018a,Grusdt2018b}. 

We consider the ground state $|\Psi_0\rangle$ at half-filling and dope a hole at position $\bm{j}^s$. The state is denoted as $|\bm{j}^s,\bar{\sigma}, 0\rangle = c_{\bm{j}^s,\sigma}|\Psi_0\rangle$. We decompose the hole into partons; a holon $h_j^\dagger$ and a spinon $f_{j,\sigma}$, i.e., $c_{\bm{j}^s,\sigma} = h_j^{\dagger} f_{j,\sigma}$. In the basis state $|\bm{j}^s,\bar{\sigma}, 0\rangle$ the spinon is therefore located at position $\bm{j}^s$ and has spin $\bar{\sigma} = -\sigma$. The motion of the holon creates different geometric strings $\Sigma$ of the displaced spins, $|\bm{j}^s,\bar{\sigma},\Sigma\rangle$, keeping the spinon position fixed. The
spinon and the holon are connected by strings of displaced spins, each of them comes at an energy cost. Thus separating the spinon and the holon leads to a potential $V_\text{pot}(\Sigma)$ that grows linearly with the length of the string $\Sigma$. In that picture of frozen spins, the holon motion can be modeled as a single particle hopping problem on the Bethe lattice with linearly increasing potential (see
End Matter for details). 
This linear dependence on string length effectively gives rise to a spinon-holon bound state of energy $E_{\text{\scriptsize{FSA}}} \sim (J/t)^{2/3}$ in agreement with the numerical results; see Fig.~\ref{fig:fig2}(a). The  bound state is given by a superposition of all strings $|\psi_{\text{\scriptsize{FSA}}}(\bm{j}^s, \bar{\sigma})\rangle =\sum_{\Sigma}\psi^{\text{\scriptsize{FSA}}}_{\Sigma}|\bm{j}^s, \bar{\sigma}, \Sigma\rangle$, and can be computed numerically by solving the Bethe lattice Hamiltonian. From that we obtain the probability distribution $p^{\text{\scriptsize{FSA}}}_{\Sigma} = |\psi^{\text{\scriptsize{FSA}}}_{\Sigma}|^2$ for a geometric string $\Sigma$. Crucially, compared to the conventional antiferromagnet, the diagonal hopping processes renormalize the probability of strings of zero length; see supplemental material~\cite{supp}.

\begin{figure}
    \centering
    \hspace*{-4mm}
    \includegraphics{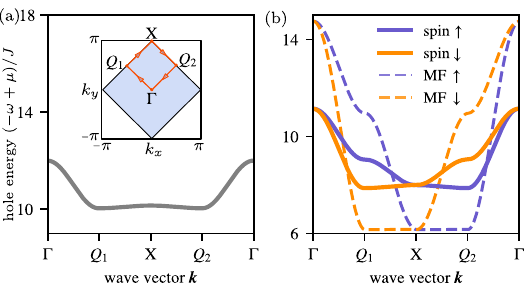}
    \caption{\textbf{Parton description of the polaron dispersion.} Predicted disperson of the magnetic polaron based on a spinon-holon parton approach (solid lines) for (a) the antiferromagnet (AFM)  with \( t_+ = 0 \) and (b) the altermagnetic Mott insulator (AMI) with \( t_+ = 0.4t \), evaluated along the Brillouin zone cut shown in the inset of (a). The AFM hosts spin-degenerate bands, while in the AMI has an anisotropic spin-splitting that governs the rotational symmetry of the $d$-wave altermagnet. Mean-field energy bands (\(U=12t, t_+=0.4t\)), obtained from Hartree-Fock as described in \cite{Das2024} (dashed lines).}
    \label{fig:fig3}
\end{figure}

The motion of the fast holon that we discussed so far assumes the position of the slow spinon to be fixed. This yields a dispersionless contribution coming from the fluctuating geometric strings. However, the spinon has a finite energy scale, and we now analyze the slow dynamics of the spinon that determines the entire dispersive part as well as the spin-split nature of the spectral function. 
For zero string length the holon and spinon are bound together, and both can hop diagonally without any energy cost, effectively contributing to the dispersive nature of the spectral function, by $E^{\text{bare}}_{t_+} (\bm{k}, \sigma)=  2t_+|\psi^{\text{\scriptsize{FSA}}}_0|^2(\cos{{k}_1}\delta_{\sigma\uparrow} + \cos{k_2}\delta_{\sigma\downarrow})$ where $k_1 = k_x +k_y$ and $k_2 = k_y -k_x$. This is the primary contribution to the spin anisotropy characteristic for altermagnetic states.

For strings $\Sigma$ of finite length, the spin-exchange coupling governs the motion of the spinon, which in terms of spinons reads $\bm{S}_i\cdot\bm{S}_j = \frac{1}{2} (f_{i\uparrow}^{\dagger}f_{i\downarrow}f_{j\downarrow}^{\dagger}f_{j\uparrow} + h.c.)+ \frac{1}{4} (n_{i\uparrow}-n_{i\downarrow})(n_{j\uparrow}-n_{j\downarrow})$. We consider all the possible spinon exchanges around the spinon at the state $|\bm{j}^s,\bar{\sigma},\Sigma\rangle$. The spinon exchanges that involve nearest neighbor sites gives a contribution $E_J(\bm{k}) = J\nu_{FC}^{(1)}[A(\cos{k_1}+\cos{k_2}) + B(\cos{2k_x}+\cos{2k_y})]$ where nearest-neighbor couplings are renormalized by the Franck-Condon factor $\nu_{FC}^{(1)} = \sum_{\text{\scriptsize{$\Sigma,\Sigma'|\Sigma$}}}\psi^{\text{\scriptsize{FSA}}}_{\Sigma}\psi^{\text{\scriptsize{FSA}}}_{\Sigma'}$, where the summation $\Sigma'|\Sigma$ is taken over all new strings $\Sigma'$ that result from the original string $\Sigma$ after a spinon exchange; the prefactors are $A = 8/3, B = 3/2$, see supplemental
material \cite{supp}. In the conventional antiferromagnetic $t-J$ model, this contribution gives a spin-degenerate dispersion \cite{Bohrdt2020,Grusdt2019}. However, in our model, spinon exchanges occur also along the diagonal for which $J_{ij} =J_+$ gives rise to $E_{J_+}(\bm{k}, \sigma) = J_+\nu_{FC}^{(2)}(\cos{k_1}\delta_{\sigma\uparrow} + \cos{k_2}\delta_{\sigma\downarrow})$, yielding a spin-split dispersion that is renormalized by Franck-Condon factor $\nu_{FC}^{(2)} = (1-|\psi^{\text{\scriptsize{FSA}}}_0|^2)$. 

In summary, the resulting momentum-dependent energy of the magnetic polaron is, 
 \begin{equation}\begin{split}
    E(\bm{k}, \sigma)&= J\nu_{FC}^{(1)}[A(\cos{k_1} + \cos{k_2}) + B(\cos{2k_x} + \cos{2k_y})] \\&
    +(J_+\nu_{FC}^{(2)}+2t_+|\psi_0^{\text{\scriptsize{FSA}}}|^2)(\cos{k_1}\delta_{\sigma\uparrow} + \cos{k_2}\delta_{\sigma\downarrow})+E_{\text{\scriptsize{FSA}}}.
\end{split}
\label{equ:tighbinding}
\end{equation}
We show the thus obtained spinon dispersion both for the antiferromagnetic and the altermagnetic case in Fig.~\ref{fig:fig3} (a) and (b), respectively. For the antiferromagent, the magnetic polaron is spin degenerate. By contrast, for the altermagnet we observe a strong spin-splitting, lifting the degeneracy of global minima at the nodal points $Q_1 = \pm(-\pi/2,\pi/2)$ ($Q_2=\pm(\pi/2,\pi/2)$) for spin $\uparrow$ ($\downarrow$). Our theory predicts that the dominant contribution to the spin-splitting arises from the rescaled free-hole contribution, $E^{\text{bare}}_{t_+}$. 

We now analyze the renormalization of the band structure due to the polaron formation in more detail. To this end, we first compute the self-consistent mean-field Hamiltonian of the altermagnetic Hubbard model in the Mott insulating regime and determine the excitation spectrum, following Ref.~\cite{Das2024}; see dashed line in \figc{fig:fig3}{b}. The correct polaron dispersion obtained from the parton theory, however, is strongly renormalized to lower energies due to the interactions with the magnetic background; solid lines in \figc{fig:fig3}{b}. This highlights the strong interaction effects, that are relevant for forming the altermagnetic polaron. 

We also compare the altermagnetic polaron dispersion obtained analytically from the spinon-holon parton theory determined with the numerical results for the spectral function; dashed lines in \figc{fig:1}{c}. 
{For a consistent comparison, we re-computed the FSA ground state energy ($E_{\text{\scriptsize{FSA}}}$) using DMRG correlations in the FSA potential energy of strings, rather than the Néel state (see supplemental material \cite{supp}). This adjustment allows us to reproduce the tensor network energy shift in our simple parton theory, resulting in excellent agreement between the numerical results and the analytical prediction for all momenta, and highlighting the importance of quantum fluctuations.}

\begin{figure}[t]
    \centering
    \hspace*{-5mm}
    \includegraphics[scale=1]{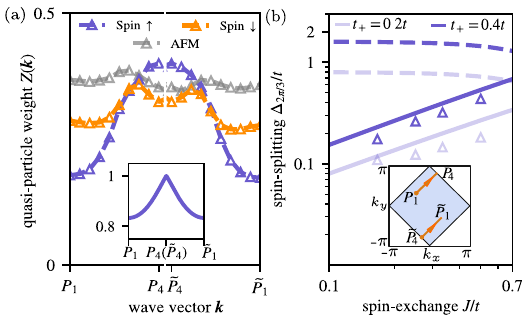}
    \caption{\textbf{Spectral signatures of altermagnets.} (a) Quasi-particle weight of the polaron branch of the electron spin-$\uparrow$ (blue) and spin-$\downarrow$ (orange) spectra for a cut through the Brillouin zone (shown in inset of b). Triangles are tensor network results. For  spin-$\uparrow$, which in the Néel ordered state is mainly affected by the diagonal hopping in the \(\bm{a}_1\) direction, the spectral weight is peaked at \(k_1 = \pi \;(P_4)\), while spin-$\downarrow$ shows a flatter spectral weight. By contrast, for this cut the antiferromagnetic response has an almost uniform quasi-particle weight. Inset: The spectral weight transfer of the altermagnet can be understood as an interference effect from hopping on our lattice geometry. (b) Spin splitting \(\Delta_{2\pi/3}\), as a function of \(J/t\) obtained from tensor networks (triangles). Self-consistent Hartree-Fock (dashed lines) strongly overestimates the spin splitting; however, the numerical results agree well with the parton theory (solid lines).}
    \label{fig:fig4}
\end{figure}

The spin-splitting of the altermagnetic polaron  leads to a transfer of spectral weight between spin polarizations; see \figc{fig:1}{c}. We determine the spectral weight  $Z(\bm{k})$  of the polaron branch at \( \bm{k} = (k_1,{2\pi}/{3}) \) by extracting the half-width at half-maximum from the spectral function, numerically obtained from tensor network simulations, see Fig. \ref{fig:fig4}(a). While for this cut in the Brillouin zone the spectral weight is approximately constant for the antiferromagnet, the situation for the altermagnet is more involved. For for spin-$\uparrow$, which for the given Néel order is mainly subjected to diagonal hopping in \( \bm{a}_1 \) direction, the spectral weight peaks at \( \bm{k} = (k_1 = \pi,{2\pi}/{3})  \) (\(P_4\)). By contrast, the weight for spin-$\downarrow$ is weakly dependent on momentum. The transfer of spectral weight results from an interference effect of the single-particle hopping on the lattice along with the spontaneous formation of Néel order, which we demonstrate by computing the quasi-particle weight of the Hartree-Fock band structure (inset of Fig.~\ref{fig:fig4}(a)).

We quantify the altermagnetic spin-splitting from the spectral function 
by taking the difference in bandwidths $\Delta_{2\pi/3} = W^A(k_1, 2\pi/3)-W^B(k_1, 2\pi/3)$ along the $(k_1,k_2=2\pi/3)$-cut of the two sublattice-resolved spectra, which we find to be numerically more stable than directly extracting the splitting at fixed momenta (see supplemental material \cite{supp}). 
The spin-splitting, measured in units of hoppings, increases as $J/t$ increases; see Fig.~\ref{fig:fig4}(b). This is understood from the limit of large $J/t$. There, the confining energy of spinon and holon increases, resulting in shorter strings on average and hence a larger renormalization factor of the free-hole contribution in Eq.~\eqref{equ:tighbinding}. Crucially, a Hartree-Fock calculation for the altermagnetic Mott insulator predicts that the splitting is weakly dependent on the ratio of $J/t$ (dashed lines), and it is severely overestimated. Hence also the spin splitting is strongly renormalized due to the formation of the altermagnetic plaron, and interaction effects are crucial to describe the ARPES spectral function.

\textbf{\textit{Conclusions and outlook.---}}We have developed a theory for the ARPES response of  altermagnetic Mott insulators. The spectrum consists of a quasi-particle branch at low energies, which we identify as the altermagnetic polaron. We find that the bandwidth of the altermagnetic polaron and the characteristic spin-splitting is strongly renormalized compared with self-consistent Hartree-Fock calculations. A key experimental signature, that we predict in addition to the spin splitting, is the spectral weight transfer between different spin polarizations; see Fig.~\ref{fig:1}(c).
For future work, the parton theory could provide an effective description of altermagnets in the effective Ising limit~\cite{Kravchuk_2025}, by extending the conventional anisotropic $t-J_z$ model~\cite{Grusdt2018a}. On a technical level, the parton theory for the polaron formation can be improved by going beyond the mean-field spinon approximation using variational techniques. Furthermore, it will be interesting to consider realistic band structures for certain insulating altermagnetic candidate materials to quantitatively compare with experiments, as well as multi-orbital altermagnetic models ~\cite{Leeb2024, Vila_2025}. 

Our theory on the altermagnetic polaron provides a systematic study of interaction effects in the ARPES response of altermagnetic Mott insulators and can be directly extended to other spectroscopic probes, such as the spin-polarized Scanning Tunneling Spectroscopy (STM) response, where a hole is locally doped into the sample. {The model we studied can be realized in cold-atomic setups as well~\cite{Das2024} and analogous information to that obtained from ARPES measurements can also be accessed such platforms~\cite{ Brown_2019,Bohrdt_2017_arpes_qgm}}. In particular, quantum gas microscopy~\cite{Bakr2009, Sherson_2010} offers a powerful probe of the spectral properties of ultracold atoms, providing complementary and, in some cases, analogous insights to ARPES in solid-state systems. Several experiments have already explored dopant effects and magnetic polaron formation in two-dimensional Fermi–Hubbard antiferromagnets~\cite{Mazurenko2017, Chiu2019, Xu_2025, prichard2025}. Our work  paves the way for a systematic study of the doping effects in altermagnetic Mott insulators, and may guide further investigations of instabilities of these states toward pair density wave and finite-momentum paired superconductors.

\textbf{\textit{Note added.---}}While preparing this manuscript, we became aware of related work studying the ARPES response of altermagnetic Mott insulators~\cite{daghofer2025}. During the review process, further related works on altermagnetic Mott insulators have appeared~\cite{Leraand_2026,park2025}.

\textbf{\textit{Acknowledgments.---}}
We thank W. Kadow, J. Knolle, and V. Leeb for insightful discussions. 
We acknowledge support from the Deutsche Forschungsgemeinschaft (DFG, German Research Foundation) under Germany’s Excellence Strategy–EXC–2111–390814868, TRR 360 – 492547816 and DFG grants No. KN1254/1-2, KN1254/2-1, the European Research Council (ERC) under the European Union’s Horizon
2020 research and innovation programme (grant agreement No 851161), the European Union (grant agreement No 101169765), as well as the Munich Quantum Valley, which is supported by the Bavarian state government with funds from the Hightech Agenda Bayern Plus.

\textbf{\textit{Data availability.---}}
Data and codes are available upon reasonable request on Zenodo~\cite{zenodo}.

\bibliography{references}

@article{Smejkal2022a,
  title = {Emerging Research Landscape of Altermagnetism},
  author = {\ifmmode \check{S}\else \v{S}\fi{}mejkal, Libor and Sinova, Jairo and Jungwirth, Tomas},
  journal = {Phys. Rev. X},
  volume = {12},
  issue = {4},
  pages = {040501},
  numpages = {27},
  year = {2022},
  month = {Dec},
  publisher = {American Physical Society},
  doi = {10.1103/PhysRevX.12.040501},
  url = {https://link.aps.org/doi/10.1103/PhysRevX.12.040501}
}

@article{Smejkal2022b,
  title = {Beyond Conventional Ferromagnetism and Antiferromagnetism: A Phase with Nonrelativistic Spin and Crystal Rotation Symmetry},
  author = {\ifmmode \check{S}\else \v{S}\fi{}mejkal, Libor and Sinova, Jairo and Jungwirth, Tomas},
  journal = {Phys. Rev. X},
  volume = {12},
  issue = {3},
  pages = {031042},
  numpages = {16},
  year = {2022},
  month = {Sep},
  publisher = {American Physical Society},
  doi = {10.1103/PhysRevX.12.031042},
  url = {https://link.aps.org/doi/10.1103/PhysRevX.12.031042}
}

@article{Smejkal2020crystal,
author = {Libor Šmejkal  and Rafael González-Hernández  and T. Jungwirth  and J. Sinova },
title = {Crystal time-reversal symmetry breaking and spontaneous Hall effect in collinear antiferromagnets},
journal = {Science Advances},
volume = {6},
number = {23},
pages = {eaaz8809},
year = {2020},
doi = {10.1126/sciadv.aaz8809},
URL = {https://www.science.org/doi/abs/10.1126/sciadv.aaz8809}}

@article{Ahn2019,
  title = {Antiferromagnetism in ${\mathrm{RuO}}_{2}$ as $d$-wave Pomeranchuk instability},
  author = {Ahn, Kyo-Hoon and Hariki, Atsushi and Lee, Kwan-Woo and Kune\ifmmode \check{s}\else \v{s}\fi{}, Jan},
  journal = {Phys. Rev. B},
  volume = {99},
  issue = {18},
  pages = {184432},
  numpages = {5},
  year = {2019},
  month = {May},
  publisher = {American Physical Society},
  doi = {10.1103/PhysRevB.99.184432},
  url = {https://link.aps.org/doi/10.1103/PhysRevB.99.184432}
}

@article{wu2007fermi,
  title = {Fermi liquid instabilities in the spin channel},
  author = {Wu, Congjun and Sun, Kai and Fradkin, Eduardo and Zhang, Shou-Cheng},
  journal = {Phys. Rev. B},
  volume = {75},
  issue = {11},
  pages = {115103},
  numpages = {25},
  year = {2007},
  month = {Mar},
  publisher = {American Physical Society},
  doi = {10.1103/PhysRevB.75.115103},
  url = {https://link.aps.org/doi/10.1103/PhysRevB.75.115103}
}

@article{kaushal2024,
      title={Altermagnetism in modified Lieb lattice Hubbard model}, 
      author={Nitin Kaushal and Marcel Franz},
      year={2024},
      journal={arXiv.2412.16421},
      url={https://arxiv.org/abs/2412.16421}, 
}

@article{Fedchenko2024,
author = {Olena Fedchenko  and Jan Minár  and Akashdeep Akashdeep  and Sunil Wilfred D’Souza  and Dmitry Vasilyev  and Olena Tkach  and Lukas Odenbreit  and Quynh Nguyen  and Dmytro Kutnyakhov  and Nils Wind  and Lukas Wenthaus  and Markus Scholz  and Kai Rossnagel  and Moritz Hoesch  and Martin Aeschlimann  and Benjamin Stadtmüller  and Mathias Kläui  and Gerd Schönhense  and Tomas Jungwirth  and Anna Birk Hellenes  and Gerhard Jakob  and Libor Šmejkal  and Jairo Sinova  and Hans-Joachim Elmers },
title = {Observation of time-reversal symmetry breaking in the band structure of altermagnetic RuO2},
journal = {Science Advances},
volume = {10},
number = {5},
pages = {eadj4883},
year = {2024},
doi = {10.1126/sciadv.adj4883},
URL = {https://www.science.org/doi/abs/10.1126/sciadv.adj4883},

}

@article{Lee2024,
  title = {Broken Kramers Degeneracy in Altermagnetic MnTe},
  author = {Lee, Suyoung and Lee, Sangjae and Jung, Saegyeol and Jung, Jiwon and Kim, Donghan and Lee, Yeonjae and Seok, Byeongjun and Kim, Jaeyoung and Park, Byeong Gyu and \ifmmode \check{S}\else \v{S}\fi{}mejkal, Libor and Kang, Chang-Jong and Kim, Changyoung},
  journal = {Phys. Rev. Lett.},
  volume = {132},
  issue = {3},
  pages = {036702},
  numpages = {7},
  year = {2024},
  month = {Jan},
  publisher = {American Physical Society},
  doi = {10.1103/PhysRevLett.132.036702},
  url = {https://link.aps.org/doi/10.1103/PhysRevLett.132.036702}
}

@Article{Amin2024,
author={Amin, O. J.
and Dal Din, A.
and Golias, E.
and Niu, Y.
and Zakharov, A.
and Fromage, S. C.
and Fields, C. J. B.
and Heywood, S. L.
and Cousins, R. B.
and Maccherozzi, F.
and Krempask{\'y}, J.
and Dil, J. H.
and Kriegner, D.
and Kiraly, B.
and Campion, R. P.
and Rushforth, A. W.
and Edmonds, K. W.
and Dhesi, S. S.
and {\v{S}}mejkal, L.
and Jungwirth, T.
and Wadley, P.},
title={Nanoscale imaging and control of altermagnetism in MnTe},
journal={Nature},
year={2024},
month={Dec},
day={01},
volume={636},
number={8042},
pages={348-353},

issn={1476-4687},
doi={10.1038/s41586-024-08234-x},
url={https://doi.org/10.1038/s41586-024-08234-x}
}

@article{Gonzalez_Betancourt2023,
  title = {Spontaneous Anomalous Hall Effect Arising from an Unconventional Compensated Magnetic Phase in a Semiconductor},
  author = {Gonzalez Betancourt, R. D. and Zub\'a\ifmmode \check{c}\else \v{c}\fi{}, J. and Gonzalez-Hernandez, R. and Geishendorf, K. and \ifmmode \check{S}\else \v{S}\fi{}ob\'a\ifmmode \check{n}\else \v{n}\fi{}, Z. and Springholz, G. and Olejn\'{\i}k, K. and \ifmmode \check{S}\else \v{S}\fi{}mejkal, L. and Sinova, J. and Jungwirth, T. and Goennenwein, S. T. B. and Thomas, A. and Reichlov\'a, H. and \ifmmode \check{Z}\else \v{Z}\fi{}elezn\'y, J. and Kriegner, D.},
  journal = {Phys. Rev. Lett.},
  volume = {130},
  issue = {3},
  pages = {036702},
  numpages = {7},
  year = {2023},
  month = {Jan},
  publisher = {American Physical Society},
  doi = {10.1103/PhysRevLett.130.036702},
  url = {https://link.aps.org/doi/10.1103/PhysRevLett.130.036702}
}

@article{Maier2023,
  title = {Weak-coupling theory of neutron scattering as a probe of altermagnetism},
  author = {Maier, Thomas A. and Okamoto, Satoshi},
  journal = {Phys. Rev. B},
  volume = {108},
  issue = {10},
  pages = {L100402},
  numpages = {5},
  year = {2023},
  month = {Sep},
  publisher = {American Physical Society},
  doi = {10.1103/PhysRevB.108.L100402},
  url = {https://link.aps.org/doi/10.1103/PhysRevB.108.L100402}
}

@article{Leeb2024,
  title = {Spontaneous Formation of Altermagnetism from Orbital Ordering},
  author = {Leeb, Valentin and Mook, Alexander and \ifmmode \check{S}\else \v{S}\fi{}mejkal, Libor and Knolle, Johannes},
  journal = {Phys. Rev. Lett.},
  volume = {132},
  issue = {23},
  pages = {236701},
  numpages = {7},
  year = {2024},
  month = {Jun},
  publisher = {American Physical Society},
  doi = {10.1103/PhysRevLett.132.236701},
  url = {https://link.aps.org/doi/10.1103/PhysRevLett.132.236701}
}

@article{gray2024,
      title={Time-resolved magneto-optical effects in the altermagnet candidate MnTe}, 
      author={Isaiah Gray and Qinwen Deng and Qi Tian and Michael Chilcote and J. Steven Dodge and Matthew Brahlek and Liang Wu},
      year={2024},
      journal={arXiv:2404.05020},
      url={https://arxiv.org/abs/2404.05020}, 
}

@article{Das2024,
  title = {Realizing Altermagnetism in Fermi-Hubbard Models with Ultracold Atoms},
  author = {Das, Purnendu and Leeb, Valentin and Knolle, Johannes and Knap, Michael},
  journal = {Phys. Rev. Lett.},
  volume = {132},
  issue = {26},
  pages = {263402},
  numpages = {7},
  year = {2024},
  month = {Jun},
  publisher = {American Physical Society},
  doi = {10.1103/PhysRevLett.132.263402},
  url = {https://link.aps.org/doi/10.1103/PhysRevLett.132.263402}
}

@article{Bey2025,
      title={Interface, bulk and surface structure of heteroepitaxial altermagnetic $\alpha$-MnTe films grown on GaAs(111)}, 
      author={Sara Bey and Maksym Zhukovskyi and Tatyana Orlova and Shelby Fields and Valeria Lauter and Haile Ambaye and Anton Ievlev and Steven P. Bennett and Xinyu Liu and Badih A. Assaf},
      year={2025},
      journal={arXiv:2504.12126},
      url={https://arxiv.org/abs/2504.12126}, 
}

@article{Ferrari2024,
  title = {Altermagnetism on the Shastry-Sutherland lattice},
  author = {Ferrari, Francesco and Valent\'{\i}, Roser},
  journal = {Phys. Rev. B},
  volume = {110},
  issue = {20},
  pages = {205140},
  numpages = {11},
  year = {2024},
  month = {Nov},
  publisher = {American Physical Society},
  doi = {10.1103/PhysRevB.110.205140},
  url = {https://link.aps.org/doi/10.1103/PhysRevB.110.205140}
}

@misc{zenodo,
  author       = {Lanzini, Lorenzo and Das, Purnendu and Knap, Michael},
  title        = {Zenodo entry for: {Theory} of {Angle} {Resolved} {Photoemission} {Spectroscopy} of {Altermagnetic} {Mott} {Insulators}},
  year         = 2025,
  publisher    = {Zenodo},
  doi          = {10.5281/zenodo.15553362},

}

@article{Krempaský2024,
author={Krempask{\'y}, J.
and {\v{S}}mejkal, L.
and D'Souza, S. W.
and Hajlaoui, M.
and Springholz, G.
and Uhl{\'i}{\v{r}}ov{\'a}, K.
and Alarab, F.
and Constantinou, P. C.
and Strocov, V.
and Usanov, D.
and Pudelko, W. R.
and Gonz{\'a}lez-Hern{\'a}ndez, R.
and Birk Hellenes, A.
and Jansa, Z.
and Reichlov{\'a}, H.
and {\v{S}}ob{\'a}{\v{n}}, Z.
and Gonzalez Betancourt, R. D.
and Wadley, P.
and Sinova, J.
and Kriegner, D.
and Min{\'a}r, J.
and Dil, J. H.
and Jungwirth, T.},
title={Altermagnetic lifting of Kramers spin degeneracy},
journal={Nature},
year={2024},
month={Feb},
day={01},
volume={626},
number={7999},
pages={517-522},
issn={1476-4687},
doi={10.1038/s41586-023-06907-7},
url={https://doi.org/10.1038/s41586-023-06907-7}
}

@article{he2025,
      title={Altermagnetism and beyond in the $t$-$t^\prime$-$\delta$ Fermi-Hubbard model}, 
      author={Saisai He and Jize Zhao and Hong-Gang Luo and Shijie Hu},
      year={2025},
      journal={arXiv:2503.08362},
      url={https://arxiv.org/abs/2503.08362}, 
}

@article{wu2025,
      title={The Fermi surface of RuO2 measured by quantum oscillations}, 
      author={Zheyu Wu and Mengmeng Long and Hanyi Chen and Shubhankar Paul and Hisakazu Matsuki and Oleksandr Zheliuk and Uli Zeitler and Gang Li and Rui Zhou and Zengwei Zhu and Dave Graf and Theodore I. Weinberger and F. Malte Grosche and Yoshiteru Maeno and Alexander G. Eaton},
      year={2025},
      journal={arXiv:2503.20621},
      url={https://arxiv.org/abs/2503.20621}, 
}

@article{vijayvargia2025,
      title={Altermagnets with topological order in Kitaev bilayers}, 
      author={Aayush Vijayvargia and Ezra Day-Roberts and Antia S. Botana and Onur Erten},
      year={2025},
      journal={arXiv:2503.09705},
      url={https://arxiv.org/abs/2503.09705}, 
}

@article{Hariki2024,
  title = {X-Ray Magnetic Circular Dichroism in Altermagnetic $\ensuremath{\alpha}$-MnTe},
  author = {Hariki, A. and Dal Din, A. and Amin, O. J. and Yamaguchi, T. and Badura, A. and Kriegner, D. and Edmonds, K. W. and Campion, R. P. and Wadley, P. and Backes, D. and Veiga, L. S. I. and Dhesi, S. S. and Springholz, G. and \ifmmode \check{S}\else \v{S}\fi{}mejkal, L. and V\'yborn\'y, K. and Jungwirth, T. and Kune\ifmmode \check{s}\else \v{s}\fi{}, J.},
  journal = {Phys. Rev. Lett.},
  volume = {132},
  issue = {17},
  pages = {176701},
  numpages = {7},
  year = {2024},
  month = {Apr},
  publisher = {American Physical Society},
  doi = {10.1103/PhysRevLett.132.176701},
  url = {https://link.aps.org/doi/10.1103/PhysRevLett.132.176701}
}

@article{Bohrdt2020,
  title = {Parton theory of angle-resolved photoemission spectroscopy spectra in antiferromagnetic Mott insulators},
  author = {Bohrdt, Annabelle and Demler, Eugene and Pollmann, Frank and Knap, Michael and Grusdt, Fabian},
  journal = {Phys. Rev. B},
  volume = {102},
  issue = {3},
  pages = {035139},
  numpages = {23},
  year = {2020},
  month = {Jul},
  publisher = {American Physical Society},
  doi = {10.1103/PhysRevB.102.035139},
  url = {https://link.aps.org/doi/10.1103/PhysRevB.102.035139}
}

@article{Sachdev1989,
  title = {Hole motion in a quantum N\'eel state},
  author = {Sachdev, Subir},
  journal = {Phys. Rev. B},
  volume = {39},
  issue = {16},
  pages = {12232--12247},
  numpages = {0},
  year = {1989},
  month = {Jun},
  publisher = {American Physical Society},
  doi = {10.1103/PhysRevB.39.12232},
  url = {https://link.aps.org/doi/10.1103/PhysRevB.39.12232}
}

@article{Grusdt2019,
  title = {Microscopic spinon-chargon theory of magnetic polarons in the $t\text{\ensuremath{-}}J$ model},
  author = {Grusdt, Fabian and Bohrdt, Annabelle and Demler, Eugene},
  journal = {Phys. Rev. B},
  volume = {99},
  issue = {22},
  pages = {224422},
  numpages = {14},
  year = {2019},
  month = {Jun},
  publisher = {American Physical Society},
  doi = {10.1103/PhysRevB.99.224422},
  url = {https://link.aps.org/doi/10.1103/PhysRevB.99.224422}
}

@article{Bohrdt_2020_b,
doi = {10.1088/1367-2630/abcfee},
url = {https://dx.doi.org/10.1088/1367-2630/abcfee},
year = {2020},
month = {dec},
publisher = {IOP Publishing},
volume = {22},
number = {12},
pages = {123023},
author = {Bohrdt, A and Grusdt, F and Knap, M},
title = {Dynamical formation of a magnetic polaron in a two-dimensional quantum antiferromagnet},
journal = {New Journal of Physics},

}

@Article{Hauschild2018,
	title={{Efficient numerical simulations with Tensor Networks: Tensor Network Python (TeNPy)}},
	author={Johannes Hauschild and Frank Pollmann},
	journal={SciPost Phys. Lect. Notes},
	pages={5},
	year={2018},
	publisher={SciPost},
	doi={10.21468/SciPostPhysLectNotes.5},
	url={https://scipost.org/10.21468/SciPostPhysLectNotes.5},
}

@article{Zaletel2015,
  title = {Time-evolving a matrix product state with long-ranged interactions},
  author = {Zaletel, Michael P. and Mong, Roger S. K. and Karrasch, Christoph and Moore, Joel E. and Pollmann, Frank},
  journal = {Phys. Rev. B},
  volume = {91},
  issue = {16},
  pages = {165112},
  numpages = {8},
  year = {2015},
  month = {Apr},
  publisher = {American Physical Society},
  doi = {10.1103/PhysRevB.91.165112},
  url = {https://link.aps.org/doi/10.1103/PhysRevB.91.165112}
}

@article{Osumi2024,
  title = {Observation of a giant band splitting in altermagnetic MnTe},
  author = {Osumi, T. and Souma, S. and Aoyama, T. and Yamauchi, K. and Honma, A. and Nakayama, K. and Takahashi, T. and Ohgushi, K. and Sato, T.},
  journal = {Phys. Rev. B},
  volume = {109},
  issue = {11},
  pages = {115102},
  numpages = {8},
  year = {2024},
  month = {Mar},
  publisher = {American Physical Society},
  doi = {10.1103/PhysRevB.109.115102},
  url = {https://link.aps.org/doi/10.1103/PhysRevB.109.115102}
}

@article{Motrunich2006,
  title = {Orbital magnetic field effects in spin liquid with spinon Fermi sea: Possible application to $\ensuremath{\kappa}\text{\ensuremath{-}}{(\mathrm{ET})}_{2}{\mathrm{Cu}}_{2}{(\mathrm{C}\mathrm{N})}_{3}$},
  author = {Motrunich, Olexei I.},
  journal = {Phys. Rev. B},
  volume = {73},
  issue = {15},
  pages = {155115},
  numpages = {11},
  year = {2006},
  month = {Apr},
  publisher = {American Physical Society},
  doi = {10.1103/PhysRevB.73.155115},
  url = {https://link.aps.org/doi/10.1103/PhysRevB.73.155115}
}

@Article{Costa2025,
	title={{Giant spatial anisotropy of magnon Landau damping in altermagnets}},
	author={António T. Costa and João C. G. Henriques and Joaquín Fernández-Rossier},
	journal={SciPost Phys.},
	volume={18},
	pages={125},
	year={2025},
	publisher={SciPost},
	doi={10.21468/SciPostPhys.18.4.125},
	url={https://scipost.org/10.21468/SciPostPhys.18.4.125},
}

@article{daghofer2025,
      title={Altermagnetic polarons}, 
      author={Maria Daghofer and Krzysztof Wohlfeld and Jeroen van den Brink},
      year={2025},
      journal={arXiv.2506.03261},
      url={https://arxiv.org/abs/2506.03261}
}

@article{Brown_2019,
   title={Angle-resolved photoemission spectroscopy of a Fermi–Hubbard system},
   volume={16},
   url={http://dx.doi.org/10.1038/s41567-019-0696-0},
   DOI={10.1038/s41567-019-0696-0},
   number={1},
   journal={Nature Physics},
   author={Brown, Peter T. and Guardado-Sanchez, Elmer and Spar, Benjamin M. and Huang, Edwin W. and Devereaux, Thomas P. and Bakr, Waseem S.},
   year={2019},
   pages={26–31} }

@article{Sherson_2010,
   title={Single-atom-resolved fluorescence imaging of an atomic Mott insulator},
   volume={467},
   url={http://dx.doi.org/10.1038/nature09378},
   DOI={10.1038/nature09378},
   number={7311},
   journal={Nature},
   author={Sherson, Jacob F. and Weitenberg, Christof and Endres, Manuel and Cheneau, Marc and Bloch, Immanuel and Kuhr, Stefan},
   year={2010},
   pages={68–72} }

@article{Xu_2025,
   title={A neutral-atom Hubbard quantum simulator in the cryogenic regime},
   volume={642},
   url={http://dx.doi.org/10.1038/s41586-025-09112-w},
   DOI={10.1038/s41586-025-09112-w},
   number={8069},
   journal={Nature},
   author={Xu, Muqing and Kendrick, Lev Haldar and Kale, Anant and Gang, Youqi and Feng, Chunhan and Zhang, Shiwei and Young, Aaron W. and Lebrat, Martin and Greiner, Markus},
   year={2025},
   pages={909–915} }

@article{Pomeranchuk1958,
  title={On the stability of a Fermi liquid},
  author={Pomeranchuk, I Ia and others},
  journal={Sov. Phys. JETP},
  volume={8},
  pages={361},
  year={1958}
}

@article{Wrzosek_2021,
   title={Hole in the two-dimensional Ising antiferromagnet: Origin of the incoherent spectrum},
   volume={103},
   ISSN={2469-9969}, 
   pages={035113},
   url={http://dx.doi.org/10.1103/PhysRevB.103.035113},
   DOI={10.1103/physrevb.103.035113},
   number={3},
   journal={Phys. Rev. B},
   author={Wrzosek, Piotr and Wohlfeld, Krzysztof},
   year={2021}
}

@article{Chiu2019,
author = {Christie S. Chiu  and Geoffrey Ji  and Annabelle Bohrdt  and Muqing Xu  and Michael Knap  and Eugene Demler  and Fabian Grusdt  and Markus Greiner  and Daniel Greif },
title = {String patterns in the doped Hubbard model},
journal = {Science},
volume = {365},
number = {6450},
pages = {251-256},
year = {2019},
URL = {https://www.science.org/doi/abs/10.1126/science.aav3587},
}

@article{Grusdt2018a,
  title = {Parton Theory of Magnetic Polarons: Mesonic Resonances and Signatures in Dynamics},
  author = {Grusdt, F. and K\'anasz-Nagy, M. and Bohrdt, A. and Chiu, C. S. and Ji, G. and Greiner, M. and Greif, D. and Demler, E.},
  journal = {Phys. Rev. X},
  volume = {8},
  issue = {1},
  pages = {011046},
  numpages = {39},
  year = {2018},
  month = {Mar},
  publisher = {American Physical Society},
  doi = {10.1103/PhysRevX.8.011046},
  url = {https://link.aps.org/doi/10.1103/PhysRevX.8.011046}
}

@Article{Grusdt2018b,
	title={{Meson formation in mixed-dimensional t-J models}},
	author={Fabian Grusdt and Zheng Zhu and Tao Shi and Eugene Demler},
	journal={SciPost Phys.},
	volume={5},
	pages={057},
	year={2018},
	publisher={SciPost},
	doi={10.21468/SciPostPhys.5.6.057},
	url={https://scipost.org/10.21468/SciPostPhys.5.6.057},
}

@misc{supp,
title = {see supplementary material.},
year = {2025}
}

@article{Sobota2021,
  title = {Angle-resolved photoemission studies of quantum materials},
  author = {Sobota, Jonathan A. and He, Yu and Shen, Zhi-Xun},
  journal = {Rev. Mod. Phys.},
  volume = {93},
  issue = {2},
  pages = {025006},
  numpages = {72},
  year = {2021},
  month = {May},
  publisher = {American Physical Society},
  doi = {10.1103/RevModPhys.93.025006},
  url = {https://link.aps.org/doi/10.1103/RevModPhys.93.025006}
}

@Article{Reimers2024,
author={Reimers, Sonka
and Odenbreit, Lukas
and {\v{S}}mejkal, Libor
and Strocov, Vladimir N.
and Constantinou, Procopios
and Hellenes, Anna B.
and Jaeschke Ubiergo, Rodrigo
and Campos, Warlley H.
and Bharadwaj, Venkata K.
and Chakraborty, Atasi
and Denneulin, Thibaud
and Shi, Wen
and Dunin-Borkowski, Rafal E.
and Das, Suvadip
and Kl{\"a}ui, Mathias
and Sinova, Jairo
and Jourdan, Martin},
title={Direct observation of altermagnetic band splitting in CrSb thin films},
journal={Nature Communications},
year={2024},
month={Mar},
day={08},
volume={15},
number={1},
pages={2116},
issn={2041-1723},
doi={10.1038/s41467-024-46476-5},
url={https://doi.org/10.1038/s41467-024-46476-5}
}

@Article{Yang2025,
author={Yang, Guowei
and Li, Zhanghuan
and Yang, Sai
and Li, Jiyuan
and Zheng, Hao
and Zhu, Weifan
and Pan, Ze
and Xu, Yifu
and Cao, Saizheng
and Zhao, Wenxuan
and Jana, Anupam
and Zhang, Jiawen
and Ye, Mao
and Song, Yu
and Hu, Lun-Hui
and Yang, Lexian
and Fujii, Jun
and Vobornik, Ivana
and Shi, Ming
and Yuan, Huiqiu
and Zhang, Yongjun
and Xu, Yuanfeng
and Liu, Yang},
title={Three-dimensional mapping of the altermagnetic spin splitting in CrSb},
journal={Nature Communications},
year={2025},
month={Feb},
day={07},
volume={16},
number={1},
pages={1442},
issn={2041-1723},
doi={10.1038/s41467-025-56647-7},
url={https://doi.org/10.1038/s41467-025-56647-7}
}

@article{Ding2024,
  title = {Large Band Splitting in $g$-Wave Altermagnet CrSb},
  author = {Ding, Jianyang and Jiang, Zhicheng and Chen, Xiuhua and Tao, Zicheng and Liu, Zhengtai and Li, Tongrui and Liu, Jishan and Sun, Jianping and Cheng, Jinguang and Liu, Jiayu and Yang, Yichen and Zhang, Runfeng and Deng, Liwei and Jing, Wenchuan and Huang, Yu and Shi, Yuming and Ye, Mao and Qiao, Shan and Wang, Yilin and Guo, Yanfeng and Feng, Donglai and Shen, Dawei},
  journal = {Phys. Rev. Lett.},
  volume = {133},
  issue = {20},
  pages = {206401},
  numpages = {7},
  year = {2024},
  month = {Nov},
  publisher = {American Physical Society},
  doi = {10.1103/PhysRevLett.133.206401},
  url = {https://link.aps.org/doi/10.1103/PhysRevLett.133.206401}
}

@article{li2024,
      title={Topological Weyl Altermagnetism in CrSb}, 
      author={Cong Li and Mengli Hu and Zhilin Li and Yang Wang and Wanyu Chen and Balasubramanian Thiagarajan and Mats Leandersson and Craig Polley and Timur Kim and Hui Liu and Cosma Fulga and Maia G. Vergniory and Oleg Janson and Oscar Tjernberg and Jeroen van den Brink},
      year={2024},
      journal={arXiv:2405.14777},
      url={https://arxiv.org/abs/2405.14777}, 
}

@article{jaeschkeubiergo2025,
      title={Atomic Altermagnetism}, 
      author={Rodrigo Jaeschke-Ubiergo and Venkata-Krishna Bharadwaj and Warlley Campos and Ricardo Zarzuela and Nikolaos Biniskos and Rafael M. Fernandes and Tomas Jungwirth and Jairo Sinova and Libor Šmejkal},
      year={2025},
      journal={arXiv:2503.10797},
      url={https://arxiv.org/abs/2503.10797}, 
}

@article{Hayami_2019,
author = {Hayami ,Satoru and Yanagi ,Yuki and Kusunose ,Hiroaki},
title = {Momentum-Dependent Spin Splitting by Collinear Antiferromagnetic Ordering},
journal = {Journal of the Physical Society of Japan},
volume = {88},
number = {12},
pages = {123702},
year = {2019},
doi = {10.7566/JPSJ.88.123702},
URL = {https://doi.org/10.7566/JPSJ.88.123702},

}

@article{mazin2021prediction,
   title={Prediction of unconventional magnetism in doped FeSb
            2},
   volume={118},
   ISSN={1091-6490},
   url={http://dx.doi.org/10.1073/pnas.2108924118},
   DOI={10.1073/pnas.2108924118},
   number={42},
   journal={Proceedings of the National Academy of Sciences},
   publisher={Proceedings of the National Academy of Sciences},
   author={Mazin, Igor I. and Koepernik, Klaus and Johannes, Michelle D. and González-Hernández, Rafael and Šmejkal, Libor},
   year={2021},
   month=oct }

@article{smejkal2022giant,
  title = {Giant and Tunneling Magnetoresistance in Unconventional Collinear Antiferromagnets with Nonrelativistic Spin-Momentum Coupling},
  author = {\ifmmode \check{S}\else \v{S}\fi{}mejkal, Libor and Hellenes, Anna Birk and Gonz\'alez-Hern\'andez, Rafael and Sinova, Jairo and Jungwirth, Tomas},
  journal = {Phys. Rev. X},
  volume = {12},
  issue = {1},
  pages = {011028},
  numpages = {11},
  year = {2022},
  month = {Feb},
  publisher = {American Physical Society},
  doi = {10.1103/PhysRevX.12.011028},
  url = {https://link.aps.org/doi/10.1103/PhysRevX.12.011028}
}

@article{Mazin2022editorial,
  title = {Editorial: Altermagnetism---A New Punch Line of Fundamental Magnetism},
  author = {Mazin, Igor},
  collaboration = {The PRX Editors},
  journal = {Phys. Rev. X},
  volume = {12},
  issue = {4},
  pages = {040002},
  numpages = {3},
  year = {2022},
  month = {Dec},
  publisher = {American Physical Society},
  doi = {10.1103/PhysRevX.12.040002},
  url = {https://link.aps.org/doi/10.1103/PhysRevX.12.040002}
}

@Article{Shao2021,
author={Shao, Ding-Fu
and Zhang, Shu-Hui
and Li, Ming
and Eom, Chang-Beom
and Tsymbal, Evgeny Y.},
title={Spin-neutral currents for spintronics},
journal={Nature Communications},
year={2021},
month={Dec},
day={03},
volume={12},
number={1},
pages={7061},
issn={2041-1723},
doi={10.1038/s41467-021-26915-3},
url={https://doi.org/10.1038/s41467-021-26915-3}
}

@article{Hayami2020,
  title = {Bottom-up design of spin-split and reshaped electronic band structures in antiferromagnets without spin-orbit coupling: Procedure on the basis of augmented multipoles},
  author = {Hayami, Satoru and Yanagi, Yuki and Kusunose, Hiroaki},
  journal = {Phys. Rev. B},
  volume = {102},
  issue = {14},
  pages = {144441},
  numpages = {24},
  year = {2020},
  month = {Oct},
  publisher = {American Physical Society},
  doi = {10.1103/PhysRevB.102.144441},
  url = {https://link.aps.org/doi/10.1103/PhysRevB.102.144441}
}

@article{white,
  title = {Density matrix formulation for quantum renormalization groups},
  author = {White, Steven R.},
  journal = {Phys. Rev. Lett.},
  volume = {69},
  issue = {19},
  pages = {2863--2866},
  numpages = {0},
  year = {1992},
  month = {Nov},
  publisher = {American Physical Society},
  doi = {10.1103/PhysRevLett.69.2863},
  url = {https://link.aps.org/doi/10.1103/PhysRevLett.69.2863}
}

@article{Schmitt-Rink1988,
  title = {Spectral Function of Holes in a Quantum Antiferromagnet},
  author = {Schmitt-Rink, S. and Varma, C. M. and Ruckenstein, A. E.},
  journal = {Phys. Rev. Lett.},
  volume = {60},
  issue = {26},
  pages = {2793--2796},
  numpages = {0},
  year = {1988},
  month = {Jun},
  publisher = {American Physical Society},
  doi = {10.1103/PhysRevLett.60.2793},
  url = {https://link.aps.org/doi/10.1103/PhysRevLett.60.2793}
}

@article{Kane1989,
  title = {Motion of a single hole in a quantum antiferromagnet},
  author = {Kane, C. L. and Lee, P. A. and Read, N.},
  journal = {Phys. Rev. B},
  volume = {39},
  issue = {10},
  pages = {6880--6897},
  numpages = {0},
  year = {1989},
  month = {Apr},
  publisher = {American Physical Society},
  doi = {10.1103/PhysRevB.39.6880},
  url = {https://link.aps.org/doi/10.1103/PhysRevB.39.6880}
}

@article{B_ran_1996,
   title={Evidence for composite nature of quasiparticles in the 2D t-J model},
   volume={473},
   ISSN={0550-3213},
   url={http://dx.doi.org/10.1016/0550-3213(96)00196-4},
   DOI={10.1016/0550-3213(96)00196-4},
   number={3},
   journal={Nuclear Physics B},
   publisher={Elsevier BV},
   author={Béran, P. and Poilblanc, D. and Laughlin, R.B.},
   year={1996},
   month=aug, pages={707–720} }

@article{Laughlin1997,
  title = {Evidence for Quasiparticle Decay in Photoemission from Underdoped Cuprates},
  author = {Laughlin, R. B.},
  journal = {Phys. Rev. Lett.},
  volume = {79},
  issue = {9},
  pages = {1726--1729},
  numpages = {0},
  year = {1997},
  month = {Sep},
  publisher = {American Physical Society},
  doi = {10.1103/PhysRevLett.79.1726},
  url = {https://link.aps.org/doi/10.1103/PhysRevLett.79.1726}
}

@article{Manousakis2007,
  title = {String excitations of a hole in a quantum antiferromagnet and photoelectron spectroscopy},
  author = {Manousakis, Efstratios},
  journal = {Phys. Rev. B},
  volume = {75},
  issue = {3},
  pages = {035106},
  numpages = {10},
  year = {2007},
  month = {Jan},
  publisher = {American Physical Society},
  doi = {10.1103/PhysRevB.75.035106},
  url = {https://link.aps.org/doi/10.1103/PhysRevB.75.035106}
}

@article{Brinkman1970,
  title = {Single-Particle Excitations in Magnetic Insulators},
  author = {Brinkman, W. F. and Rice, T. M.},
  journal = {Phys. Rev. B},
  volume = {2},
  issue = {5},
  pages = {1324--1338},
  numpages = {0},
  year = {1970},
  month = {Sep},
  publisher = {American Physical Society},
  doi = {10.1103/PhysRevB.2.1324},
  url = {https://link.aps.org/doi/10.1103/PhysRevB.2.1324}
}

@article{GUO_2023,
title = {Spin-split collinear antiferromagnets: A large-scale ab-initio study},
journal = {Materials Today Physics},
volume = {32},
pages = {100991},
year = {2023},
issn = {2542-5293},
doi = {https://doi.org/10.1016/j.mtphys.2023.100991},
url = {https://www.sciencedirect.com/science/article/pii/S2542529323000275},
author = {Yaqian Guo and Hui Liu and Oleg Janson and Ion Cosma Fulga and Jeroen {van den Brink} and Jorge I. Facio},

}

@article{Maznichenko_2024,
  title = {Fragile altermagnetism and orbital disorder in Mott insulator ${\mathrm{LaTiO}}_{3}$},
  author = {Maznichenko, I. V. and Ernst, A. and Maryenko, D. and Dugaev, V. K. and Sherman, E. Ya. and Buczek, P. and Parkin, S. S. P. and Ostanin, S.},
  journal = {Phys. Rev. Mater.},
  volume = {8},
  issue = {6},
  pages = {064403},
  numpages = {6},
  year = {2024},
  month = {Jun},
  publisher = {American Physical Society},
  doi = {10.1103/PhysRevMaterials.8.064403},
  url = {https://link.aps.org/doi/10.1103/PhysRevMaterials.8.064403}
}

@article{Cuono_2023,
title = {Orbital-selective altermagnetism and correlation-enhanced spin-splitting in strongly-correlated transition metal oxides},
journal = {Journal of Magnetism and Magnetic Materials},
volume = {586},
pages = {171163},
year = {2023},
issn = {0304-8853},
doi = {https://doi.org/10.1016/j.jmmm.2023.171163},
url = {https://www.sciencedirect.com/science/article/pii/S0304885323008132},
author = {Giuseppe Cuono and Raghottam M. Sattigeri and Jan Skolimowski and Carmine Autieri},

}

@article{Wei_2025,
  title = {${\mathrm{La}}_{2}{\mathrm{O}}_{3}{\mathrm{Mn}}_{2}{\mathrm{Se}}_{2}$: A correlated insulating layered d-wave altermagnet},
  author = {Wei, Chao-Chun and Li, Xiaoyin and Hatt, Sabrina and Huai, Xudong and Liu, Jue and Singh, Birender and Kim, Kyung-Mo and Fernandes, Rafael M. and Cardon, Paul and Zhao, Liuyan and Tran, Thao T. and Frandsen, Benjamin A. and Burch, Kenneth S. and Liu, Feng and Ji, Huiwen},
  journal = {Phys. Rev. Mater.},
  volume = {9},
  issue = {2},
  pages = {024402},
  numpages = {13},
  year = {2025},
  month = {Feb},
  publisher = {American Physical Society},
  doi = {10.1103/PhysRevMaterials.9.024402},
  url = {https://link.aps.org/doi/10.1103/PhysRevMaterials.9.024402}
}

@Article{Mazurenko2017,
author={Mazurenko, Anton
and Chiu, Christie S.
and Ji, Geoffrey
and Parsons, Maxwell F.
and Kan{\'a}sz-Nagy, M{\'a}rton
and Schmidt, Richard
and Grusdt, Fabian
and Demler, Eugene
and Greif, Daniel
and Greiner, Markus},
title={A cold-atom Fermi--Hubbard antiferromagnet},
journal={Nature},
year={2017},
month={May},
day={01},
volume={545},
number={7655},
pages={462-466},
issn={1476-4687},
doi={10.1038/nature22362},
url={https://doi.org/10.1038/nature22362}
}

@article{prichard2025,
      title={Observation of Magnon-Polarons in the Fermi-Hubbard Model}, 
      author={Max L. Prichard and Zengli Ba and Ivan Morera and Benjamin M. Spar and David A. Huse and Eugene Demler and Waseem S. Bakr},
      year={2025},
      journal={arXiv.2502.06757},
      url={https://arxiv.org/abs/2502.06757}
}

@article{Bohrdt_2017_arpes_qgm,
  title = {Angle-resolved photoemission spectroscopy with quantum gas microscopes},
  author = {Bohrdt, A. and Greif, D. and Demler, E. and Knap, M. and Grusdt, F.},
  journal = {Phys. Rev. B},
  volume = {97},
  issue = {12},
  pages = {125117},
  numpages = {23},
  year = {2018},
  month = {Mar},
  publisher = {American Physical Society},
  doi = {10.1103/PhysRevB.97.125117},
  url = {https://link.aps.org/doi/10.1103/PhysRevB.97.125117}
}

@Article{Bakr2009,
author={Bakr, Waseem S.
and Gillen, Jonathon I.
and Peng, Amy
and F{\"o}lling, Simon
and Greiner, Markus},
title={A quantum gas microscope for detecting single atoms in a Hubbard-regime optical lattice},
journal={Nature},
year={2009},
month={Nov},
day={01},
volume={462},
number={7269},
pages={74-77},

issn={1476-4687},
doi={10.1038/nature08482},
url={https://doi.org/10.1038/nature08482}
}

@article{Chang_2025,
  title = {Energy dispersion, superconductivity, and magnetic fluctuations in stacked altermagnetic materials},
  author = {Chang, Jun and Lu, Hantao and Zhao, Jize and Luo, Hong-Gang and Ding, Yang},
  journal = {Phys. Rev. B},
  volume = {111},
  issue = {10},
  pages = {104432},
  numpages = {8},
  year = {2025},
  month = {Mar},
  publisher = {American Physical Society},
  doi = {10.1103/PhysRevB.111.104432},
  url = {https://link.aps.org/doi/10.1103/PhysRevB.111.104432}
}

@article{DelRe_2025,
  title = {Dirac points and topological phases in correlated altermagnets},
  author = {Del Re, Lorenzo},
  journal = {Phys. Rev. Res.},
  volume = {7},
  issue = {3},
  pages = {033234},
  numpages = {11},
  year = {2025},
  month = {Sep},
  publisher = {American Physical Society},
  doi = {10.1103/7nvm-s225},
  url = {https://link.aps.org/doi/10.1103/7nvm-s225}
}

@article{Parshukov2025,
  author       = {Kirill Parshukov and Raymond Wiedmann and Andreas P. Schnyder},
  title        = {Topological crossings in two-dimensional altermagnets: Symmetry classification and topological responses},
  journal      = {Physical Review B},
  volume       = {111},
  pages        = {224406},
  year         = {2025},
  month        = {Jun},
  doi          = {10.1103/PhysRevB.111.224406}
}

@article{Sato2024,
  author       = {Toshihiro Sato and Sonia Haddad and Ion Cosma Fulga and Fakher F. Assaad and Jeroen van den Brink},
  title        = {Altermagnetic Anomalous Hall Effect Emerging from Electronic Correlations},
  journal      = {Physical Review Letters},
  volume       = {133},
  number       = {8},
  pages        = {086503},
  year         = {2024},
  month        = {Aug},
  doi          = {10.1103/PhysRevLett.133.086503}
}

@article{Liu_2025,
  title = {Quantum dynamics in a spin-$\frac{1}{2}$ square lattice ${J}_{1}\text{\ensuremath{-}}{J}_{2}\text{\ensuremath{-}}\ensuremath{\delta}$ altermagnet},
  author = {Liu, Yang and Shao, Shiqi and He, Saisai and Xie, Z. Y. and Mei, Jia-Wei and Luo, Hong-Gang and Zhao, Jize},
  journal = {Phys. Rev. B},
  volume = {111},
  issue = {24},
  pages = {245117},
  numpages = {7},
  year = {2025},
  month = {Jun},
  publisher = {American Physical Society},
  doi = {10.1103/PhysRevB.111.245117},
  url = {https://link.aps.org/doi/10.1103/PhysRevB.111.245117}
}

@article{StrackVollhardt1992,
  author       = {Rainer Strack and Dieter Vollhardt},
  title        = {Dynamics of a hole in the t-J model with local disorder: exact results for high dimensions},
  journal      = {Physical Review B},
  volume       = {46},
  number       = {22},
  pages        = {13852--13861},
  year         = {1992},
  month        = {Dec},
  doi          = {10.1103/PhysRevB.46.13852}
}

@article{LoganStumpf1998,
  author       = {D. E. Logan and M. P. H. Stumpf},
  title        = {Finite-temperature hole dynamics in the t-J model: Exact results for high dimensions},
  journal      = {Europhysics Letters},
  volume       = {43},
  number       = {2},
  pages        = {207--212},
  year         = {1998},
  doi          = {10.1209/epl/i1998-00341-6}
}

@article{Sangiovanni2006,
  author       = {G. Sangiovanni and A. Toschi and K. Held and M. Capone and O. Gunnarsson and S.-K. Mo and J. W. Allen},
  title        = {Static versus dynamical mean-field theory of Mott antiferromagnets},
  journal      = {Physical Review B},
  volume       = {73},
  pages        = {205121},
  year         = {2006},
  doi          = {10.1103/PhysRevB.73.205121}
}

@article{Taranto2012,
  author       = {C. Taranto and S. Andergassen and J. Bauer and A. Kauch and K. Held and G. Sangiovanni and A. Toschi},
  title        = {Signature of antiferromagnetic long-range order in the optical conductivity of correlated systems},
  journal      = {Physical Review B},
  volume       = {85},
  pages        = {085124},
  year         = {2012},
  doi          = {10.1103/PhysRevB.85.085124}
}

@Article{Regmi2025,
author={Regmi, Resham Babu
and Bhandari, Hari
and Thapa, Bishal
and Hao, Yiqing
and Sharma, Nileema
and McKenzie, James
and Chen, Xinglong
and Nayak, Abhijeet
and El Gazzah, Mohamed
and M{\'a}rkus, Bence G.
and Forr{\'o}, L{\'a}szl{\'o}
and Liu, Xiaolong
and Cao, Huibo
and Mitchell, J. F.
and Mazin, Igor I.
and Ghimire, Nirmal J.},
title={Altermagnetism in the layered intercalated transition metal dichalcogenide CoNb4Se8},
journal={Nature Communications},
year={2025},
month={May},
day={13},
volume={16},
number={1},
pages={4399},
issn={2041-1723},
doi={10.1038/s41467-025-58642-4},
url={https://doi.org/10.1038/s41467-025-58642-4}
}

@article{devita2025,
      title={Optical switching in a layered altermagnet}, 
      author={Alessandro De Vita and Chiara Bigi and Davide Romanin and Matthew D. Watson and Vincent Polewczyk and Marta Zonno and François Bertran and My Bang Petersen and Federico Motti and Giovanni Vinai and Manuel Tuniz and Federico Cilento and Mario Cuoco and Brian M. Andersen and Andreas Kreisel and Luciano Jacopo D'Onofrio and Oliver J. Clark and Mark T. Edmonds and Christopher Candelora and Muxian Xu and Siyu Cheng and Alexander LaFleur and Tommaso Antonelli and Giorgio Sangiovanni and Lorenzo Del Re and Ivana Vobornik and Jun Fujii and Fabio Miletto Granozio and Alessia Sambri and Emiliano Di Gennaro and Jeppe B. Jacobsen and Henrik Jacobsen and Ralph Ernstorfer and Ilija Zeljkovic and Younghun Hwang and Matteo Calandra and Jill A. Miwa and Federico Mazzola},
      year={2025},
      journal={arXiv.2502.20010},
      url={https://arxiv.org/abs/2502.20010}, 
}

@article{gong2025,
      title={Tunability of the magnetic properties in Ni intercalated transition metal dichalcogenide NbSe$_2$}, 
      author={Xujia Gong and Amar Fakhredine and Carmine Autieri},
      year={2025},
      journal={arXiv.2505.17916},
      url={https://arxiv.org/abs/2505.17916}, 
}

@article{Kadow_2022,
   title={Hole spectral function of a chiral spin liquid in the triangular lattice Hubbard model},
   volume={106},
   ISSN={2469-9969},
   url={http://dx.doi.org/10.1103/PhysRevB.106.094417},
   DOI={10.1103/physrevb.106.094417},
   number={9},
   journal={Physical Review B},
   publisher={American Physical Society (APS)},
   author={Kadow, Wilhelm and Vanderstraeten, Laurens and Knap, Michael},
   year={2022},
   month=sep }

@article{Zhu2025,
  title = {Design of Altermagnetic Models from Spin Clusters},
  author = {Zhu, Xingchuan and Huo, Xingmin and Feng, Shiping and Zhang, Song-Bo and Yang, Shengyuan A. and Guo, Huaiming},
  journal = {Phys. Rev. Lett.},
  volume = {134},
  issue = {16},
  pages = {166701},
  numpages = {6},
  year = {2025},
  month = {Apr},
  publisher = {American Physical Society},
  doi = {10.1103/PhysRevLett.134.166701},
  url = {https://link.aps.org/doi/10.1103/PhysRevLett.134.166701}
}

@book{auerbach1998interacting,
  title={Interacting electrons and quantum magnetism},
  author={Auerbach, Assa},
  year={1998},
  publisher={Springer Science \& Business Media}
}

@article{wiedmann2025,
      title={Quantum effects in the magnon spectrum of 2D altermagnets via continuous similarity transformations}, 
      author={Raymond Wiedmann and Dag-Björn Hering and Vanessa Sulaiman and Matthias R. Walther and Kai P. Schmidt and Götz S. Uhrig},
      year={2025},
      journal={arXiv.2511.03528},
      url={https://arxiv.org/abs/2511.03528}, 
}

@article{peces2025,
      title={Altermagnetism in an interacting model of Kagome materials}, 
      author={Alejandro Blanco Peces and Jaime Merino},
      year={2025},
      journal={arXiv.2510.21291},
      url={https://arxiv.org/abs/2510.21291}, 
}

@article{sicheler2025,
      title={Optically Tunable Spin Transport in Bilayer Altermagnetic Mott Insulators}, 
      author={Niklas Sicheler and Roberto Raimondi and Giorgio Sangiovanni and Lorenzo Del Re},
      year={2025},
      journal={arXiv.2508.06938},
      url={https://arxiv.org/abs/2508.06938}, 
}

@article{ni2025,
      title={Competitive Orders in Altermagnetic Chiral Magnons}, 
      author={Congzhe Yan Zhijun Jiang Jinyang Ni and Guoqing Chang},
      year={2025},
      journal={arXiv.2511.03922},
      url={https://arxiv.org/abs/2511.03922}, 
}

@article{Kravchuk_2025,
   title={Chiral magnetic excitations and domain textures of 
   g-wave altermagnets},
   volume={112},
   ISSN={2469-9969},
   url={http://dx.doi.org/10.1103/zn8d-ft9b},
   DOI={10.1103/zn8d-ft9b},
   number={14},
   journal={Physical Review B},
   publisher={American Physical Society (APS)},
   author={Kravchuk, Volodymyr P. and Yershov, Kostiantyn V. and Facio, Jorge I. and Guo, Yaqian and Janson, Oleg and Gomonay, Olena and Sinova, Jairo and van den Brink, Jeroen},
   year={2025},
   month=Oct }

@article{Leraand_2026,
  title = {Spin-dependent quasiparticle lifetimes in altermagnets},
  author = {Leraand, Kristoffer and M\ae{}land, Kristian and Sudb\o{}, Asle},
  journal = {Phys. Rev. B},
  volume = {113},
  issue = {11},
  pages = {115148},
  numpages = {17},
  year = {2026},
  month = {Mar},
  publisher = {American Physical Society},
  doi = {10.1103/r3vm-m1m3},
  url = {https://link.aps.org/doi/10.1103/r3vm-m1m3}
}

@article{park2025,
      title={Impact of strong electronic correlations on altermagnets: the case of NiS2}, 
      author={Ina Park and Turan Birol and Antoine Georges and Rafael M. Fernandes},
      year={2025},
     journal={arXiv.2512.17059},
      url={https://arxiv.org/abs/2512.17059}, 
}

@article{Vila_2025,
  title = {Orbital-spin locking and its optical signatures in altermagnets},
  author = {Vila, Marc and Sunko, Veronika and Moore, Joel E.},
  journal = {Phys. Rev. B},
  volume = {112},
  issue = {2},
  pages = {L020401},
  numpages = {8},
  year = {2025},
  month = {Jul},
  publisher = {American Physical Society},
  doi = {10.1103/bzzy-ngcs},
  url = {https://link.aps.org/doi/10.1103/bzzy-ngcs}
}

@Article{Garcia-Gassull2026,
author={Garcia-Gassull, Laura
and Razpopov, Aleksandar
and Stavropoulos, P. Peter
and Mazin, Igor I.
and Valent{\'i}, Roser},
title={Microscopic origin of the magnetic interactions and their experimental signatures in altermagnetic La2O3Mn2Se2},
journal={npj Spintronics},
year={2026},
month={Feb},
day={26},
volume={4},
number={1},
pages={9},

issn={2948-2119},
doi={10.1038/s44306-025-00125-9},
url={https://doi.org/10.1038/s44306-025-00125-9}
}
\newpage
\subsection{End Matter}
\subsubsection{Geometric string theory and Frozen Spin Approximation}\label{supp:FSA th}
\begin{figure}[ht]
    \centering
    \hspace*{-7mm}
    \includegraphics[scale=1]{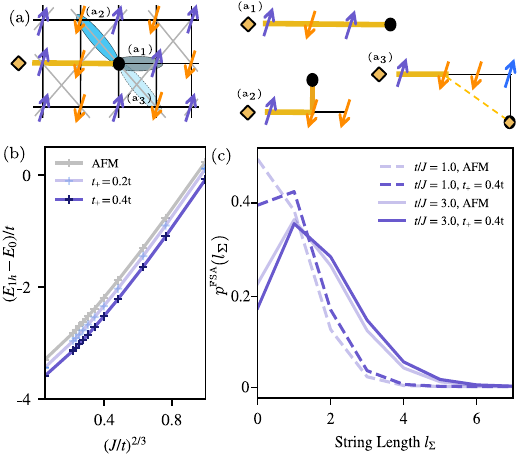}
    \caption{\textbf{Spinon-holon parton theory.}  \textbf{(a)} We illustrate possible holon hopping paths that lead to different spin configurations: the nearest-neighbor hopping \( t \)  \( (\text{a}_1) \), which increases the string length by one unit, the diagonal hopping \( t_+ \) that couples strings of the same length \( (\text{a}_2) \), or breaks the string and creates a second excitation \( (\text{a}_3) \).  \textbf{(b)} The energy difference for the state with injected hole and the one at half filling scales approximately with $(J/t)^{2/3}$, predicted by the analytical solutions of the single-particle Schrodinger equation. \textbf{(c)} String length distribution for $t/J=1$ (dotted lines), $t/J=3$ (solid lines) and $t_+=0$ and $t_+=0.4t$. With increasing hopping strength, the distribution broadens.}
    \label{fig:app 1}
\end{figure}

We describe the dynamics of a single hole in the altermagntic $t-J$ model using a spinon-holon parton description of the magnetic polaron~\cite{B_ran_1996, Laughlin1997,Grusdt2018a, Grusdt2019, Bohrdt2020, Bohrdt_2020_b}. The hole in the altermagnetic background can be described as a bound state of a heavy spinon carrying a spin quantum number and the light holon carrying the charge. The partons are interconnected by geometric strings which we analyze based on the Frozen Spin Approximation (FSA)~\cite{Grusdt2018a, Grusdt2018b} where we assume that the motion of the holon does not alter the quantum state of the spins and only displaces them along the string; see Fig.~\ref{fig:app 1}(a). 

The electron annihilation operator can be written as
\begin{equation}
    c_{j,\sigma} = h_j^{\dagger} f_{j,\sigma}
\end{equation}
where $h_j^{\dagger}$ is the bosonic holon creation operator and $f_{j,\sigma}$ is the fermionic spinon operator.
The physical Hilbert space is composed of all states satisfying
\begin{equation}
    \sum_{\sigma} f_{j,\sigma}^{\dagger} f_{j,\sigma}+ h_j^{\dagger}h_j = 1 \hspace{0.5cm}\forall j.
\end{equation}
First, we consider the ground state $|\Psi_0\rangle$ of the $t-J$ model of the altermagnet at half-filling. We create a hole at $\bm{j}^s$ by applying $c_{\bm{j}^s,\sigma}$. The spinon is located at $\bm{j}^s$ with spin $\bar{\sigma}$. Hence, we get a state with a spinon and a holon
\begin{equation}
    |\bm{j}^s,\bar{\sigma},0\rangle = c_{\bm{j}^s,\sigma}|\Psi_0\rangle = h_{\bm{j}^s}^{\dagger} f_{\bm{j}^s,\sigma}|\Psi_0\rangle
\end{equation}
At strong coupling ($t\gg J$, i.e., $U\gg t$), we assume the FSA by considering the spinon to be fixed at $\bm{j}^s$ and determine the fast holon motion. To this end, we can decompose the Hilbert space as, 
    $\mathcal{H} = \mathcal{H}_{\Sigma} \otimes \mathcal{H}_\text{spinon}$.
Here, $\mathcal{H}_\text{spinon}$ is the space spanned by all the spinon states and $\mathcal{H}_{\Sigma}$ is the space spanned by all possible strings of the holon departing from the spinon. The state after creating a string $\Sigma$ is given, by
\begin{equation}
    |\bm{j}^s,\bar{\sigma},\Sigma\rangle = G_{\Sigma}h_{\bm{j}^s}^{\dagger} f_{\bm{j}^s,\sigma}|\Psi_0\rangle
\end{equation}
with the string operator
\begin{equation}
    G_{\Sigma} = \prod_{\langle i,j\rangle \in \Sigma} (h_i^{\dagger}h_j\sum_{\tau =\uparrow,\downarrow}f_{j,\tau}^{\dagger}f_{i,\tau})
\end{equation}
where $\langle i,j\rangle$ is along the string $\Sigma$. Most of the states $|\bm{j}^s,\bar{\sigma},\Sigma\rangle$, which are created in a classical Néel background are orthonormal to each other \cite{Bohrdt2020,Grusdt2019}. An exception are so-called Trugman loops, i.e., strings where the holon goes back to its initial position, that change the holon position without additional strings. However, by a counting argument, it can be shown that their effect on the spinon dispersion is small \cite{Grusdt2019}. Further corrections can be obtained by considering self-avoiding walks on the square lattice, which we do not take into account for simplicity~\cite{Wrzosek_2021}. With these approximations, the Hilbert space created by the states $|\bm{j}^s,\bar{\sigma},\Sigma\rangle$ has the form of the Bethe lattice (as in the antiferromagnetic case \cite{Grusdt2019}), but now some of the nodes on different branches are interconnected due to the presence of diagonal hopping.
In this space, the effective Hamiltonian becomes, $H_\text{eff} = H_{\Sigma} + H_\text{pot} + H_\text{spinon}$, where the hopping on the sites of the Bethe lattice gives
\begin{equation}
    H_{\Sigma} = -\sum_{\bm{j}^s,\sigma}\sum_{\langle\Sigma_i,\Sigma_j\rangle} t_{ij}(|\bm{j}^s,\bar{\sigma},\Sigma_i\rangle\langle\bm{j}^s,\bar{\sigma},\Sigma_j|+h.c.).
\end{equation}
Here, for the strings $\Sigma_i,\Sigma_j$ that are separated by nearest neighbors $t_{ij}$ is $t$, and for strings $\Sigma_i,\Sigma_j$ that are separated by a diagonal $t_{ij}$ is $t_+~(0)$ in the $(1,1)$ direction and $0~(t_+)$ in the $(-1,1)$ direction for the A (B) sublattice. The motion of the hole creates the string of displaced spins that gives rise to the potential energy given by,
\begin{equation}
    H_\text{pot} = \sum_{\bm{j}^s,\sigma}\sum_{\Sigma} \langle\bm{j}^s,\bar{\sigma},\Sigma|H_{J}|\bm{j}^s,\bar{\sigma},\Sigma\rangle|\bm{j}^s,\bar{\sigma},\Sigma\rangle\langle\bm{j}^s,\bar{\sigma},\Sigma| 
\end{equation}
 The tight-binding Hamiltonian of the spinon, $H_\text{spinon}$, also appears in the effective Hamiltonian, which involves the exchange of heavy spinons, which we will analyze in the next section.
 We determine the potential energy $V_\text{pot}(\Sigma) = \langle\bm{j}^s,\bar{\sigma},\Sigma|H_{J}|\bm{j}^s,\bar{\sigma},\Sigma\rangle$ and its dependence on the string length $l_{\Sigma}$ by looking at the correlations among the spins. In a linear approximation, we determine the potential energy as
 \begin{equation}
     V_\text{pot}(\Sigma) = \frac{dE}{dl} l_{\Sigma} + g_0\delta_{l_{\Sigma},0}+ g_1 \delta_{l_{\Sigma},1} + E_0
 \end{equation}
with $g_0 = -J(C_2-C_1), g_1 = -J_+(C_2-C_1)/2, dE/dl = 2J(C_2-C_1)+ J_+(C_1/2+ C_4 - 3C_2/2), E_0 =  J(1+C_2 - 5C_1) + J_+(1/2-2C_2)$ where the correlations $C_d = \langle \Psi_0 | \bm{S}_d \cdot \bm{S}_0|\Psi_0\rangle$ are determined from the undoped ground state $|\Psi_0\rangle$.{Here, the sublabel "d" indicates the ordered distance from the other spin we take the correlation with (nearest-neighbor $d=1$, next-nearest-neighbor (diagonal) $d=2$, etc.).} 

\begin{figure}
    \centering
    \includegraphics[scale=0.54]{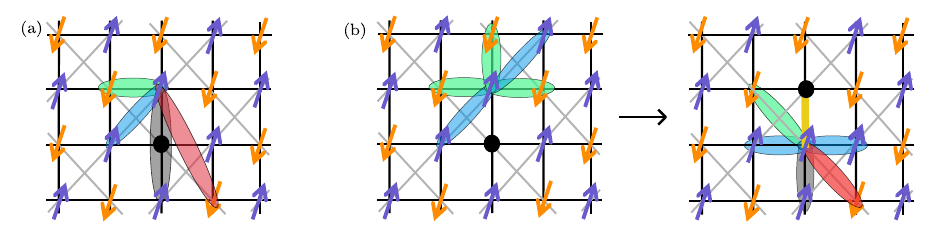}
    \caption{{(a) We highlight NN (C1), N-NN (C2), N-N-NN (C3) and N-N-N-NN (C4) correlations with green, cyan, grey and red color respectively. In the perfect Néel state, $C_1=C_4=-1/4$ and $C_2=C_3=1/4$. (b) The energy cost of a single hole (left) is $E_{l_{\Sigma}=0}-E_{AFM} = -4JC_1-2J_+ C_2 + J +J_+/2 = -J(4C_1-1)-J_+(2C_2 -1/2)= E_0 + g_0$. The energy cost of $l_{\Sigma}=1$ (right) is $E_{l_{\Sigma}=1}-E_{AFM}= J(1+3C_2-7C_1) + J_+(1/2-4C_2+C_4+C_1) = E_0 + g_1 + dE/dl$.}}
    \label{fig:linear potential l1}
\end{figure}

\begin{figure}
    \centering
    \includegraphics[scale=0.75]{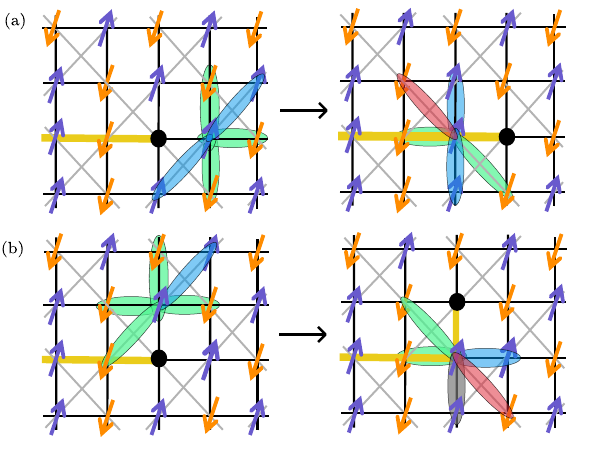}
    \caption{{The holon hops and increases the string length; to calculate $\tfrac{dE}{dl}$, we average between two slightly different cases: (a) the new spin of the string was not interacting with the string before the hopping and (b) the new spin of the string was interacting via a $J_+$-term with the string. In (a), we get $E_{l_{\Sigma}+1}-E_{l_{\Sigma}}= J(2C_2+C_1-3C_1) + J_+(C_4+C_1-2C_2) = 2J(C_2-C_1) + J_+(C_1-2C_2 + C_4)$. In (b), we get $E_{l_{\Sigma}+1}-E_{l_{\Sigma}}= J(C_1+C_3+C_2-3C_1) + J_+(C_4+C_1-C_2-C_1) = 2J(C_2-C_1) + J_+(C_4-C_2)$. We average between these two cases and obtain $dE/dl = 2J(C_2-C_1)+ J_+(C_1/2+ C_4 - 3C_2/2)$.}}
    \label{fig:linear potential}
\end{figure}

{We now motivate the coefficients of the potential energy of strings; first, with our coefficients the $l_{\Sigma} = 0$ case gives
\begin{equation}\begin{aligned}
    V_{\text{pot}}&(l_{\Sigma}=0) = g_0 + E_0 \\ &=  -J(C_2-C_1) + J(1+C_2 - 5C_1) + J_+(1/2-2C_2) \\&= -J(4C_1 - 1) -J_+(2C_2 -1/2).
\end{aligned}\end{equation}
This result can be checked by calculating the energy cost of removing a particle, see Fig.\ref{fig:linear potential l1}. We also notice that, when expanding on the perfect Neel state, we have $C_2=C_3=1/4$ and $C_1=C_4=-1/4$; in this approximation, we get $V_{\text{pot}}(l_{\Sigma}=0) = 2J$. 
We also determine the potential energy of string with length 1:
\begin{equation}
    \begin{aligned}
    V_{\text{pot}}&(l_{\Sigma}=1) = E_{l_{\Sigma}=1} - E_{AFM} \\ &= J(1+7C_1+2C_2+C_3)+J_+(1/2+C_1+C_4-4C_2)\\
    &= E_0 + g_1 + dE/dl \\
    &=\tfrac{7J}{2} -J_+
    \end{aligned}
    \end{equation}
(fig. \ref{fig:linear potential l1}(b)), where the last equality holds for the Néel state. To conclude, we determine the dependence of the potential energy on the string length, $dE/dl$, by analyzing long-string patterns and taking the difference $E_{l_{\Sigma}+1}-E_{l_{\Sigma}}$; we show our procedure in Fig.\ref{fig:linear potential}.} 

 We numerically determine the  ground state (spinon-holon bound state) of the Hamiltonian $H_\text{pot}+ H_{\Sigma}$, which is superposition of the strings with probabilities $p^{FSA}_{\Sigma} = |\psi^{FSA}_{\Sigma}|^2$, see Fig. \ref{fig:app 1}(c).
\begin{equation}
    |\psi_\text{FSA}(\bm{j}^s, \bar{\sigma})\rangle =\sum_{\Sigma}\psi^{FSA}_{\Sigma}|\bm{j}^s, \bar{\sigma}, \Sigma\rangle\:. \label{equ:spinon-holon bound state}
\end{equation}
 The linear dependence of the potential on the length of the string leads to the scaling of $E_\text{FSA}\sim(J/t)^{2/3}$, which is shown in Fig.~\ref{fig:app 1}(b), consistent with the tensor network simulations. 

The Hamiltonian $H_\text{pot}+H_{\Sigma}$  considers the spinon to be fixed, whereas $H_\text{spinon}$ includes the terms that involve the motion of heavy spinons. At $l_{\Sigma} = 0$, the diagonal hopping does not alter the Néel background, i.e., the potential energy stays the same. This leads to a free-hole behavior in the Néel ordered state, where the spinon and the holon always stay together. We include this contribution in $H_\text{spinon}$ (see Supplemental Material \cite{supp}), which gives a dispersive contribution to the spectrum.

\newpage
\newpage
\leavevmode \newpage

\setcounter{equation}{0}
\setcounter{page}{1}
\setcounter{figure}{0}
\renewcommand{\thepage}{S\arabic{page}}  
\renewcommand{\thefigure}{S\arabic{figure}}
\renewcommand{\theequation}{S\arabic{equation}}
\onecolumngrid
\begin{center}
\textbf{Supplemental Material:}\\
\textbf{\papertitle}\\ \vspace{10pt}

Lorenzo Lanzini$^{1, 2}$, Purnendu Das $^{3, 1, 2}$, and Michael Knap $^{1, 2}$ \\  \vspace{6pt}

$^1$\textit{\small{Technical University of Munich, TUM School of Natural Sciences, Physics Department, 85748 Garching, Germany}} \\
$^2$\textit{\small{Munich Center for Quantum Science and Technology (MCQST), Schellingstr. 4, 80799 M{\"u}nchen, Germany}} \\
$^3$\textit{\small{Indian Institute of Science, Bangalore, 560012, India}}\\
\vspace{10pt}
\end{center}
\maketitle
\twocolumngrid

\subsection{Tight-binding description of spinons}\label{supp: tb spinon}
\begin{figure}[t]
    \centering
    \hspace*{-4mm}
    \includegraphics[scale=1]{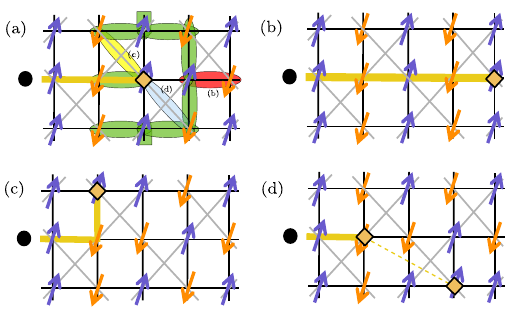}
    \caption{\textbf{Spinon Dynamics.} Off-diagonal exchanges of the Heisenberg Hamiltonian are responsible of the slow spinon dynamics. Here, we show the simplified tight-binding picture we follow to calculate the spinon dispersion. We start from a string configuration (a) and we see that nearest-neighbor exchanges $J$ (green) couple string configuration with $\Delta l_{\Sigma} = \pm 2$. Among them, we present the specific case highlighted in red (b). We show two cases of diagonal exchanges $J_+$, one coupling strings of same lengths (yellow, c) and one breaking the string and creating a second excitation (sky-blue, d).}
    \label{fig:app 2}
\end{figure}
We analyze the slow spinon dynamics with an approximated tight-binding description, by considering the spinon exchange terms, that give rise to the dispersive nature of the spectrum. We start out with the spin-interaction
\begin{equation}
\bm{S}_i\cdot\bm{S}_j = \frac{1}{2} (f_{i\uparrow}^{\dagger}f_{i\downarrow}f_{j\downarrow}^{\dagger}f_{j\uparrow} +h.c.) + \frac{1}{4} (n_{i\uparrow}-n_{i\downarrow})(n_{j\uparrow}-n_{j\downarrow}).\label{equ:spin-exchange}
\end{equation}
We can see that the first term in Eq.~\eqref{equ:spin-exchange} exchanges an up spin at site $j$ and a down spin at site $i$. 
All possible spin exchanges around the spinon at $\bm{j}^s$ in the state $|\bm{j}^s, \bar{\sigma}, \Sigma\rangle$ are shown in Fig.~\ref{fig:app 2}(a). The spin exchanges that includes $\langle ij\rangle$ to be nearest neighbors are shown with a green oval shape in Fig.~\ref{fig:app 2}(a). These exchanges change strings with length $l_{\Sigma}$ into strings of length $l_{\Sigma}\pm2$. We also illustrate an example of a nearest neighbor spin exchange using a red oval (Fig.~\ref{fig:app 2}(b)) that results into a new string configuration extended along $x$ direction with a string of length $l_{\Sigma}+2$. These nearest-neighbor spin exchanges result in a shift of the spinon position by any of the vectors $\pm 2 \hat{x}$, $ \pm2 \hat{y}$, $\pm\bm{a}_1$ or $\pm\bm{a}_2$. Addressing these nearest-neighbor $\langle ij \rangle$ exchanges, the dispersive contribution is given by
\begin{equation}
    E_J(\bm{k}) = J\nu_{FC}^{(1)}(A(\cos{k_1}+\cos{k_2}) + B(\cos{2k_x}+\cos{2k_y}))
\end{equation}
where $k_1 = k_x+k_y, k_2 = k_y - k_x$, $A=8/3$ and $B=3/2$. The value of the coefficient $A(B)$ is determined by averaging over all possible nearest-neighbor spin exchanges and string configurations that shift the spinon position by $\pm\bm{a}_1$ or $\pm\bm{a}_2$ ($\pm 2 \hat{x}$ or $ \pm2 \hat{y}$). The presence of the geometric strings renormalizes the coupling $J$ by Franck-Condon factor $\nu_{FC}^{(1)} = \sum_{\text{\scriptsize{$\Sigma,\Sigma'|\Sigma$}}}\psi^{FSA}_{\Sigma}\psi^{FSA}_{\Sigma'}$ where the summation $\Sigma'|\Sigma$ is taken over all new strings $\Sigma'$ that result from the original string $\Sigma$ after a spinon exchange. The coefficients $\psi^{\text{FSA}}_{\Sigma}$ correspond to the FSA amplitudes derived from Eq.~\eqref{equ:spinon-holon bound state}. This contribution gives a spin-degenerate dispersion that is observed in the conventional antiferromagnets as well~\cite{Grusdt2019, Bohrdt2020}.

In our altermagnetic model, we have additional couplings along the diagonal for which we have non-zero $J_{ij} = J_+$. 
The spin exchanges that includes $ij$ to be along these diagonals are shown with a yellow and sky-blue oval shape in Fig.\ref{fig:app 2}(a).
Here, we distinguish between two types of spin exchanges. In the first case (yellow), the spin exchange shifts the spinon along the diagonal that results in a new string configuration with the same string length (Fig. \ref{fig:app 2}(c)). The contribution of these exchanges is spin-dependent. If we have an up spinon, exchanges are only allowed along the $\bm{a}_2$ direction, whereas for a down spinon, exchanges occur only along the $\bm{a}_1$ direction. In the second case (shown in sky-blue oval in Fig. \ref{fig:app 2}(a)), the spin exchange breaks the existing string and separates the spinon from the string that connects to the holon (Fig. \ref{fig:app 2}(d)). We neglect the latter case as they give rise to vanishing overlaps in the trial wave function.
These diagonal next-nearest neighbor $ij$ exchanges give a contribution
\begin{equation}
    E_{J_+}(\bm{k}, \sigma) = J_+\nu_{FC}^{(2)}(\cos{k_1}\delta_{\sigma\uparrow} + \cos{k_2}\delta_{\sigma\downarrow}),
\end{equation}
where $\sigma$ is the spin of the removed electron, i.e., we have spinon with $\bar{\sigma}$. 
Once again, the presence of the geometric strings renormalizes the coupling $J_+$ by the Franck-Condon factor $\nu_{FC}^{(2)} = (1-|\psi^{FSA}_0|^2)$. The two terms $E_{J_+}(\bm{k}, \sigma)$ and $E_J(\bm{k})$ will contribute to $H_\text{spinon}$.

As discussed in the last section, at zero string length ($l_{\Sigma} = 0$), the holon and spinon are bound together. Hence, spinon and holon can hop together along the diagonal direction with hopping $t_+$. This is what we refer to as the free-hole contribution, which effectively gives rise to the dispersive nature of the spectral function. This contribution also gives rise to spin-split dispersion and is found to be the primary source of the spin splitting. If we have an up (down) spinon, then the spinon can only move along the diagonal in $\bm{a}_2\ (\bm{a}_1)$ direction. 
This contribution is given by
\begin{equation}
   E^{\text{bare}}_{t_+} (\bm{k}, \sigma)=  2t_+|\psi^{FSA}_0|^2(\cos{k_1}\delta_{\sigma\uparrow} + \cos{k_2}\delta_{\sigma\downarrow}).
\end{equation}
The hopping $t_+$ is normalized by the probability of getting a string of length $l_{\Sigma} = 0$, i.e., $|\psi^{FSA}_0|^2$. $E^{\text{bare}}_{t_+}(\bm{k}, \sigma)$ gives a stronger spin splitting compared to $E_{J_+}(\bm{k}, \sigma)$.
Summing up all these contributions and adding the total spinon-holon binding energy $E_\text{FSA}$, we get the resulting momentum-dependent energy of the magnetic polaron which is given by,
\begin{equation}
     E(\bm{k}, \sigma) = E_J(\bm{k})+E_{J_+}(\bm{k}, \sigma)+E^{\text{bare}}_{t_+}(\bm{k}, \sigma)+E_\text{FSA},
\end{equation}
which is reproducing Eq.~\eqref{equ:tighbinding} in the main text. {To consistently compare this momentum-dependent polaron energy to the MPS spectra, we recompute $E_{FSA}$ by inserting DMRG spin-spin correlations in the potential energy of strings. In particular, at $t=3J, t_+/t=0.4$, we get $C_1 \approx -0.12$, $C_2 \approx C_3 \approx 0.08, C_4 \approx -0.07$. With the new FSA energy, we obtain the excellent agreement of the two methods shown in Fig. \ref{fig:1}.}
\begin{figure}[t]
    \centering
    
    \includegraphics[scale=1]{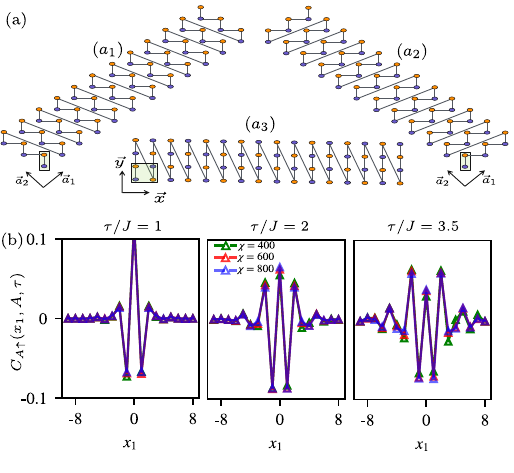}
    \caption{\textbf{Details of numerical MPS simulations.} (a) Three different cylinder geometries used: long-$\bm{a}_{1(2)}$ $(a_{1(2)})$ with two-site unit cell and long-$\bm{x} ~(a_3)$ cylinder with enlarged four-site unit cell. (b) Correlations on sublattice A along the $\bm{a}_1$ direction at different times $\tau/ J$ with $t/J=4, t_+/t=0.2$ for various bond dimensions to check convergence.}
    \label{fig:app 3}
\end{figure}

\subsection{Numerical Tensor Network Simulations}\label{supp:num details}
We perform numerical MPS simulations of our model on elongated cylinders. We consider two different cylinder geometries elongated in the \(\bm{a}_1=\bm{x}+\bm{y}\) and \(\bm{a}_2=\bm{y}-\bm{x}\) directions, respectively. Each of these cylinders consists of unit cells with two sites. We put the unit cells arranged around the circumference and eighteen along the axis, resulting in a total of \(3 \times 18 \times 2 = 108\) sites. Additionally, we choose a third cylinder geometry that is elongated in the \(\bm{x}\) direction. This cylinder features four-site unit cells, yielding a configuration of \(2 \times 14 \times 4 = 112\) sites. A visual representation of these geometries is shown in Fig.~\ref{fig:app 3} (a).

The spectral function is calculated as the Fourier transform of the time-dependent correlation functions
\begin{equation}\begin{split}
    C^{a}_{\sigma}(\bm{r}, t) &= \sum_{\text{\tiny{$b = \{A, B\}$}}}\bra{\psi_0}e^{i\mathcal{H}t}c_{\bm{r}lb\sigma}^{\dagger}e^{-i\mathcal{H}t}c_{0 a\sigma}\ket{\psi_0} \\
    &= \sum_{\text{\tiny{$b = \{A, B\}$}}}\bra{\psi_0}c_{\bm{r}b\sigma}^{\dagger}e^{-i\mathcal{H}t}c_{0 a\sigma}\ket{\psi_0}e^{i\mathcal{E}_0t}
\end{split}\end{equation}
where we have used the fact that $\ket{\psi_0}$ is the ground state of the Hamiltonian with energy $\mathcal{E}_0$. We proceed as follows:
\begin{itemize}
    \item Determine the ground state without a hole $\ket{\psi_0}$, using Density Matrix Renormalization Group (DMRG) \cite{white};
    \item Perform the time evolution of the ground state after a hole was
     created at the origin $\ket{\psi(t)}=e^{-i\mathcal{H}t}c_{0,a,\sigma}\ket{\psi_0}$ with the MPO method \cite{Zaletel2015};
     \item Calculate the overlap of $\ket{\psi(t)}$ with the state $e^{i\mathcal{E}_0t}c_{\bm{r}b\sigma}\ket{\psi_0}$, where $\mathcal{E}_0$ is the  ground state energy at half filling obtained from DMRG. 
\end{itemize}

\begin{figure}
    \centering
    \hspace*{-5mm}
    \includegraphics[scale=1]{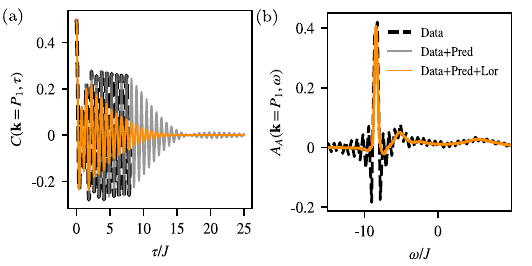}
    \caption{\textbf{Linear Prediction and Lorentzian weight.} Effect of linear prediction and Lorentzian weight on time-dependent correlations (a) and on the spectral function (b) at momentum $\mathbf{k}=P_1$.}
    \label{fig:app 4}
\end{figure}

The accuracy of the simulations is controlled by the bond dimension $\chi$ of the MPS. The time evolutions are performed up to times $5J <\tau_{max} < 8J$ with a time step $\delta\tau =0.05J$. We compare the correlation functions obtained with different $\chi = 400, 600, 800$ in Fig.~\ref{fig:app 3}(c) at different real times $\tau$. Convergence is achieved for the largest two bond dimensions shown. 

Once the real space-time correlations are obtained, we calculate the Fourier transform:
\begin{equation}
    C_{\sigma}({\mathbf{k}, \tau}) = \sum_{\bm{r}, a,b}e^{-i\mathbf{k}\cdot\bm{r}-i(b-a)k_y}C^a_\sigma(\bm{r}, \tau)
\end{equation}
In order to address the Gibbs phenomenon we use linear prediction and a filter to reduce the impact of finite time; see Fig. \ref{fig:app 4}.

With the modified correlations, we then calculate the Fourier transform in time and obtain the spectral function:
\begin{equation}
    A_{\sigma}(\mathbf{k}, \omega) = \tfrac{1}{2\pi}\int _{-\infty}^{+\infty} d\tau e^{i\tau\omega}C_{\sigma}(\mathbf{k}, \tau)
\end{equation} 
where negative-time correlations are obtained by complex conjugation of corresponding positive-time ones. We observe that linear prediction combined with Lorentzian weight reduces the impact of Gibbs oscillations and the presence of unphysical $A_{\sigma}(\mathbf{k}, \omega) < 0$; see Fig.\ref{fig:app 4}(b). {To better compare with experiments, we shift our spectra by the half filling chemical potential $\mu_{hf}$. In the antiferromagnetic model, $\mu_{hf} = U/2$ due to the particle-hole symmetry on the bipartite lattice. In our case, we compute $\mu_{hf}$ by taking the difference of ground state energies with $N+1$ and $N-1$ particles, $\mu_{hf}= \tfrac{E(N+1)-E(N-1)}{2}$. $E(N+1)$ and $E(N-1)$ are obtained from DMRG simulations of the altermagnetic Fermi Hubbard model by constraining the number of particles. For $t_+/t=0.4$ and $U=12t \: (t=3J)$, we obtain $\mu_{hf} \approx 17.4J = 5.8t \lesssim U/2$. }

\begin{figure}
    \centering
    \hspace*{-3mm}
    \includegraphics[scale=1]{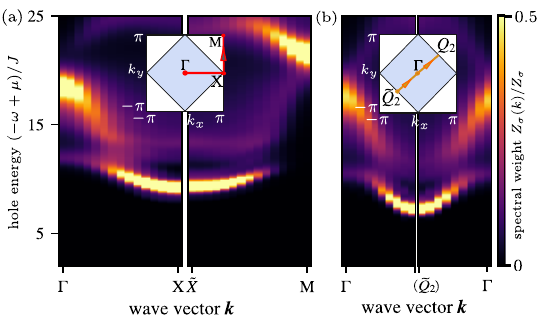}
    \caption{\textbf{Additional cuts in the Brillouin Zone.} Single-hole spectral function obtained for $t/J=3$ and $t_+/t=0.4$ on the cylinder elongated in the $\bm{x}$-direction. Weakly dispersive behaviour and suppression of spectral weight at $M=(\pi, \pi)$ are observed as in the antiferromagnetic spectrum \cite{Bohrdt2020}. (b) Diagonal cut in the Brillouin zone obtained with cylinder elongated in the $\bm{a}_1$-direction shows the global minimum at the nodal point $Q_2=(\pi/2, \pi/2).$}
    \label{fig:app 5}
\end{figure}

In Fig.\ref{fig:app 5}, we show the single-hole spectral function of the altermagnetic model along additional cuts in the Brillouin zone showing the minimum at the nodal point $Q_2=(\pi/2, \pi/2)$. 

To obtain the spin-resolved spectra of Fig.\ref{fig:1}(c), we add a small staggering field $\sum_{\bm{r}\sigma}\tfrac{-(-1)^{\sigma}B}{2}(n_{\bm{r}A\sigma}-n_{\bm{r}B\sigma})$ to the Hamiltonian ($B/J = 5\cdot10^{-3}$). This way, we pin  a finite staggered magnetization $m = \tfrac{1}{4N_u}\sum_{\bm{r}}(n_{\bm{rA\uparrow}}-n_{\bm{rA\downarrow}}-n_{\bm{rB\uparrow}}+n_{\bm{rB\downarrow}})$. With $t/J=3$ and $t_+/t=0.4$, we obtain $m \simeq 0.25$, i.e. half of the saturation value of the perfect Néel order. 

In the main text, we also mentioned the sublattice-resolved spectral function, which is obtained by choosing the sublattice where the hole is initially doped and without the need of additional staggering field:
\begin{equation}
    A^{a}(\mathbf{k}, \omega) = \tfrac{1}{2\pi}\sum_{\sigma}\int _{-\infty}^{+\infty} d\tau e^{i\tau\omega}C^{a}_{\sigma}(\mathbf{k}, \tau).
\end{equation} 
Sublattice-resolved spectra can act as a proxy for spin-resolved spectra, as each sublattice corresponds to a specific spin species. This method provides a clearer distinction between the two spin branches, as the not-fully saturated Néel order and the peak broadening can modify the peak positions. We then quantify the spin-splitting with 
\begin{equation}\begin{split}
    \Delta_{2\pi/3} &= W^{A}(k_1, 2\pi/3) - W^{B}(k_1, 2\pi/3)\\ &= W^{A}(k_1, 2\pi/3) - W^{A}(2\pi/3, k_2)\:\:, 
\end{split}\end{equation}
where in the second equality we used the altermagnetic symmetry that gives rise to the spin-anisotropy. 

\subsection{Spin-Wave theory of the Heisenberg Altermagnet}
{Here, we briefly report the spin-wave theory of the altermagnetic Heisenberg model. We re-write the altermagnetic Heisenberg Hamiltonian
\begin{equation}
\begin{aligned}
H_{JJ_+} = &\sum_{\langle ij\rangle, \sigma} J\,\vec{S}_i \cdot \vec{S}_j \\
\quad + &\sum_{(mn), \sigma} J_+\,\vec{S}_m \cdot \vec{S}_n
\end{aligned}
\end{equation}
with Holstein–Primakoff (HP) transformations \cite{auerbach1998interacting}, which map spin operators to bosonic operators distinguishing the two sublattices: 
\begin{equation}\label{hp transformations (a)}
\left\{
\begin{aligned}
    &S_{iA}^z = S -a_i^{\dagger}a_i  = 1/2 - a_i^{\dagger}a_i\\
    &S_{iA}^- = a_i^{\dagger}\sqrt{1-\tfrac{a_i^{\dagger}a_i}{2S}} \:\: \eqsim \sqrt{2S}a_i^{\dagger} =a_i^{\dagger}\\
    &S_{iA}^+ = \sqrt{1-\tfrac{a_i^{\dagger}a_i}{2S}}\:a_i \:\: \eqsim \sqrt{2S}a_i =a_1
\end{aligned}
\right.
\end{equation}
\begin{equation}\label{hp transformations (b)}
\left\{
\begin{aligned}
&S_{iB}^z = - S + b_i^{\dagger}b_i = - 1/2 + b_i^{\dagger}b_i \\
    &S_{iB}^- =\sqrt{1-\tfrac{b_i^{\dagger}b_i}{2S}}\:b_i  \:\: \eqsim \sqrt{2S}b_i =b_i \\
    &S_{iB}^+ = b_i ^{\dagger}\sqrt{1-\tfrac{b_i^{\dagger}b_i}{2S}} \:\: \eqsim \sqrt{2S}b_i^{\dagger} = b_i^{\dagger}
\end{aligned}
\right.
\end{equation}
Here, we have assumed $ \langle a_i^{\dagger}a_i\rangle /\text{\small{S}} \ll 1, \langle b_i^{\dagger}b_i\rangle /\text{\small{S}} \ll 1 $, which is formally true only in the large-S limit. To find an analytic solution, we neglect quartic order terms in the Hamiltonian (linear spin-wave theory). We then find:
\begin{equation}
    H_{JJ_+} = \sum_{q} \Psi_{\mathbf{q}}^{\dagger}h_{\mathbf{q}} \Psi_{\mathbf{q}}
\end{equation}
where
\begin{equation}
\Psi_{\mathbf{q}} = \begin{bmatrix}
    a_{\mathbf{q}}\\
    b_{\mathbf{q}}^{\dagger}
\end{bmatrix}
\end{equation}
and 
\begin{equation}\label{hq matrix}
h_{\mathbf{q}} = \begin{bmatrix}
    2J-J_+ + J_+\cos{q_1} & &  J(\cos{q_x}+\cos{q_y})\\
    J(\cos{q_x}+\cos{q_y}) & &   2J-J_+ + J_+\cos{q_2}
\end{bmatrix} 
\end{equation}
To diagonalize the 2x2 matrix, we perform a Bogoulioubov transformation:
\begin{equation}
    \begin{aligned}
    & a_q = \cosh{\theta_\mathbf{q}}\alpha_\mathbf{q} - \sinh{\theta_\mathbf{q}}\beta_\mathbf{q}^{\dagger} \\
    & b_q = -\sinh{\theta_\mathbf{q}}\alpha_{\mathbf{q}}^{\dagger} + \cosh{\theta_\mathbf{q}}\beta_\mathbf{q}
    \end{aligned}
\end{equation}
Where the choice of hyperbolic functions ensures correct bosonic commutation relations of $\alpha$ and $\beta$ operators. Rewriting Eq. \ref{hq matrix} in terms of the new bosons, we find that the condition
\begin{equation}
    \tanh{2\theta_\mathbf{q}} = \frac{\overbrace{J(\cos{q_x}+\cos{q_y})}^{\gamma(\mathbf{q})}}{\underbrace{(2J-J_+) + \tfrac{J_+}{2}(\cos{q_1}+\cos{q_2})}_{R(\mathbf{q})}}
\end{equation}
make the 2x2 matrix diagonal with eigenvalues $\omega_\mathbf{q}^{\alpha}$ and $\omega_\mathbf{q}^{\beta}$:
\begin{equation}
    \begin{aligned}t
    & \omega_{\mathbf{q}}^{\alpha} = R(\mathbf{q})\sqrt{\tfrac{R^2(\mathbf{q})}{R^2(\mathbf{q})-\gamma^2(\mathbf{q})}} -\gamma(\mathbf{q})\sqrt{\tfrac{R^2(\mathbf{q})}{R^2(\mathbf{q})-\gamma^2(\mathbf{q})}} + J_+(\cos{q_1}-\cos{q_2})\\
    & \omega_\mathbf{q}^{\beta} = R(\mathbf{q})\sqrt{\tfrac{R^2(\mathbf{q})}{R^2(\mathbf{q})-\gamma^2(\mathbf{q})}} -\gamma(\mathbf{q})\sqrt{\tfrac{R^2(\mathbf{q})}{R^2(\mathbf{q})-\gamma^2(\mathbf{q})}}  + J_+(\cos{q_2}-\cos{q_1})\: .
    \end{aligned}
\end{equation}
With the diagonalized Hamiltonian, we can define the ground state $\ket{\Omega}$ as the state without bosonic quasiparticles:
\begin{equation}
    \alpha_\mathbf{q}\ket{\Omega} = \beta_\mathbf{q}\ket{\Omega} = 0\:\: .
\end{equation}
We then calculate the staggered magnetization as 
\begin{equation}\begin{aligned}
    m &=(\langle \Omega| S^z_{iA} - S^z_{iB}|\Omega\rangle)/2 \\ &= \tfrac{1}{2} -\tfrac{1}{2N_u}\sum_{\mathbf{q}}\langle\Omega|a_\mathbf{q}^{\dagger}a_\mathbf{q} - b_\mathbf{q}^{\dagger}b_\mathbf{q}|\Omega\rangle \:\: \\
    & = \tfrac{1}{2} - \tfrac{1}{2N_u}\sum_{\mathbf{q}} \sinh^2{\theta_\mathbf{q}} \:\: .
\end{aligned}\end{equation}
We evaluate this sum numerically. With $J_+=0$, we recover the antiferromagnetic spin-wave magnetization $m_{\text{\scriptsize{AFM}}} \approx 0.304$, while for $J_+/J = 0.4^2$ we obtain $m \approx 0.279$.} 
\end{document}